\documentclass[%
reprint,
amsmath,amssymb,
aps,
prb,
twocolumn,
floatfix,
]{revtex4-2}

\usepackage{graphicx}
\usepackage{dcolumn}
\usepackage{bm}
\usepackage{stmaryrd}
\usepackage{pxfonts}
\usepackage{txfonts}
\usepackage{float}
\usepackage{ulem}
\usepackage{amssymb}
\usepackage{amsmath}
\usepackage[dvipsnames]{xcolor}
\usepackage{textcomp}
\usepackage{color}
\usepackage{multirow}
\usepackage{afterpage}

\begin{document}

\title{Quasi-Dirac points in electron-energy spectra of crystals}

\author{Grigorii\ P.\ Mikitik}
\affiliation{B. Verkin Institute for Low Temperature Physics and Engineering of the National Academy of Sciences of Ukraine, Kharkiv 61103, Ukraine}


\begin{abstract}
Specific properties, such as surface Fermi arcs, features of quantum oscillations and of various responses to a magnetic field, distinguish Dirac semimetals from ordinary materials.
These properties are determined by Dirac points at which a contact of two electron-energy bands occurs and in the vicinity of which these bands disperse linearly in the quasimomentum. This work shows that almost the same properties are inherent in a wider class of materials in which the Dirac spectrum can have a noticeable gap comparable with the Fermi energy. In other words, the degeneracy of the bands at the point and their linear dispersion are not necessary for the existence of these  properties. The only sufficient condition is the following: In the vicinity of such a quasi-Dirac point, the two close bands are well described by a two-band model that takes into account the strong spin-orbit interaction. To illustrate the results, the spectrum of ZrTe$_5$ is considered. This spectrum  contains a special quasi-Dirac point, similar to that in  bismuth.
\end{abstract}

\maketitle

\section{Introduction}

In recent years much attention has been given to the so-called Dirac semimetals; see, e.g., reviews \cite{armit,gao} and references therein. In these semimetals, two electron energy bands contact at discrete (Dirac) points of the Brillouin zone and disperse linearly in all directions around these  nodes. The Dirac points can exist in centrosymmetric crystals with a strong spin-orbit (SO) interaction. All the bands of such crystals are double degenerate in spin, and the Dirac points can have the following positions in the Brillouin zone:
(i) Points with time reversal invariant momenta (TRIM) when some variable control parameter $x$ is equal to a specific value $x_0$ \cite{murakami}. This case is realized, e.g., in the  Bi$_{1-x}$Sb$_x$ alloys at the point L of their Brillouin zone, with the parameter $x$ being the concentration of Sb, and $x_0\approx 0.04$ \cite{ponomarev}. (ii) Two symmetrically located points in a  $3$, $4$, or $6$-fold rotation axis when $x$ lies in a certain interval of its values  \cite{yang}. Such a pair of the points is created by the progressive inversion of the two bands in the axis, and the gap between the inverted bands in the middle of the axis can be considered as the parameter $x$. These pairs of the Dirac points were experimentally discovered in Na$_3$Bi \cite{liu1} and Cd$_3$As$_2$  \cite{liu,neupane}. (iii) TRIM on the Brillouin-zone surfaces of  non-symmorphic crystals when the symmetry-enforced degeneracy of the bands occurs at such points \cite{yang}.

The Dirac points determine special properties of the Dirac semimetals. In particular, the band inversion that produces the couple of the Dirac points also leads to appearance of the surface Fermi arcs \cite{yang}. Such arcs were observed in Na$_3$Bi, using the angle-resolved photoemission spectroscopy \cite{xu}. Besides, the phase of quantum oscillations are widely used to detect the Dirac fermions (see numerous references, e.g., in review article \cite{m-sh19}) since this phase is noticeably affected  by the Berry curvature produced by the Dirac points. The Dirac spectrum also manifests itself in the magneto-optical conductivity \cite{chen,yuan}, the magnetic susceptibility and the magnetic torque \cite{Moll,Zhang,modic18,m-sh19}, the magnetostriction \cite{cichorek},  and in the temperature correction to the quantum-oscillation frequency \cite{guo,alexandr}.

However, in contrast to the Weyl nodes, the Dirac points have zero Chern number \cite{armit}, and so they are not protected by this topological invariant. Therefore, a small gap can appear in the Dirac spectrum under a little variation of the crystal potential. In particular, this gap appears when the above-mentioned parameter $x$ slightly deviates from the value $x_0$, or if a uniaxial stress decreases the symmetry of the rotation axis, in which the two Dirac points lie. (The Dirac points cannot occur in the $2$-fold symmetry axis \cite{yang}). Moreover, even without any external stress, if the magnetic field is not aligned with the axis of the two Dirac points, the magnetostriction of the crystal leads to its deformation. This  deformation lifts the band degeneracy at the points. Thus, the existence of a small gap in the Dirac spectrum must be kept in mind when analyzing various experiments with the Dirac semimetals. Besides, even if the Dirac points are absent in a crystal, two bands can approach each other in a rotation axis without their crossing (see Results). If the gap between these bands is essentially smaller than the energy spacings separating them from the other bands in this region of the Brillouin zone, the electron energy spectrum can be described by a two-band ${\bf k}\cdot{\bf p}$ model.  In crystals with strong SO interaction, this model reduces to the tilted Dirac spectrum with the gap.
These situations for which the crossing of the two approaching bands is forbidden by the rotation symmetry are no less common than the case of the ordinary Dirac nodes.
We will call all such gapped spectra the quasi-Dirac spectra and  the point at which the direct gap reaches its minimal value will be named  the quasi-Dirac (QD) point. Hence, the QD points are a generalization of the usual Dirac nodes.

In this article, we draw  attention to the fact that the quasi-Dirac and Dirac spectra lead to almost identical physical properties of  crystals.
It is necessary to emphasize that this statement, which is intuitively evident for the case of a small gap as compared to the Fermi energy reckoned from the edge of one of the two close bands, is true even if the gap is essentially larger than this Fermi level, and therefore if the dispersion of the bands is not linear near the Fermi energy.
We discuss in detail conditions under which this QD spectrum can appear in real crystals. The properties of the quasi-Dirac and Dirac points distinguish the charge carriers near these points from the carriers with other types of their spectra. It is also worth noting that for a QD point at a time reversal invariant momentum, the quasi-Dirac spectrum with a variable gap can describe the topological  transition from an ordinary  insulator to the topological one. As an example, we analyze the spectrum of ZrTe$_5$, in which a special quasi-Dirac point occurs. The specificity of this point is due to the layering of this material. Using published experimental data on the Shubnikov-de Haas oscillations in ZrTe$_5$, a simple (minimal) model of its spectrum is proposed, and the model parameters are estimated.

\section{Results}

{\bf The QD spectrum.}
Consider the dispersion of two charge-carrier bands $c$ and $v$ in the vicinity of a point where they approach each other. This dispersion follows from a $4\times 4$ ${\bf k}\cdot{\bf p}$ Hamiltonian and can always be reduced to the form \cite{m-sh19}:
\begin{eqnarray}\label{1}
\varepsilon_{c,v}({\bf p})=\varepsilon_{\rm D}+{\bf a}{\bf p}\pm \sqrt{\Delta^2+v_{x}^2p_x^2 +v_{y}^2p_y^2+v_{z}^2p_z^2}.
\end{eqnarray}
Here the quasimomentum ${\bf p}$ is measured from the point (the QD point) where the direct gap in the spectrum is minimal, $2\Delta$ is the value of this minimal gap, the constants $v_i$ and ${\bf a}=(a_x,a_y,a_z)$ are the matrix elements of the velocity operator. At $\Delta=0$, formula (\ref{1}) describes the case of the Dirac spectrum, and $\varepsilon_{\rm D}$ is the energy of the Dirac point. When ${\bf a}=0$, equation (\ref{1}) is similar to the well-known Dirac equation for relativistic particles, with  $\Delta$ playing the role of the mass term \cite{LL4}. The parameter ${\bf a}$ specifies the so-called tilt of the Dirac spectrum. Such a tilt is absent for real relativistic particles, but for the pair of the Dirac points lying in the rotation axis, the vector ${\bf a}$ is aligned with this axis and differs from zero. At a nonzero ${\bf a}$, the minimal gap in spectrum (\ref{1}) is not direct, and it is equal to $2|\Delta_{\rm min}|$ (Fig.~1) where $|\Delta_{\rm min}|=|\Delta|(1-\tilde a^2)$, $\tilde a^2= \tilde a_x^2 + \tilde a_y^2 +\tilde a_z^2$, and $\tilde a_i\equiv a_i/v_i$. Below we  always assume that the rotation axis coincides with the $z$ axis, i.e., ${\bf a}=(0,0,a_z)$, and we also imply that $\tilde a^2<1$ (this condition means that the Dirac point is of the type I \cite{chang17}). The Fermi energy $\varepsilon_{\rm F}$ may have an arbitrary position relative to the edges $\varepsilon_{\rm D}\pm \Delta_{\rm min}$ of the bands, and if $\varepsilon_{\rm F}-\varepsilon_{\rm D} \sim \pm \Delta_{\rm min}$, the dispersion strongly deviates from the linear law.

\begin{figure}[tbp] 
 \centering  \vspace{+9 pt}
 \includegraphics[scale=0.41]{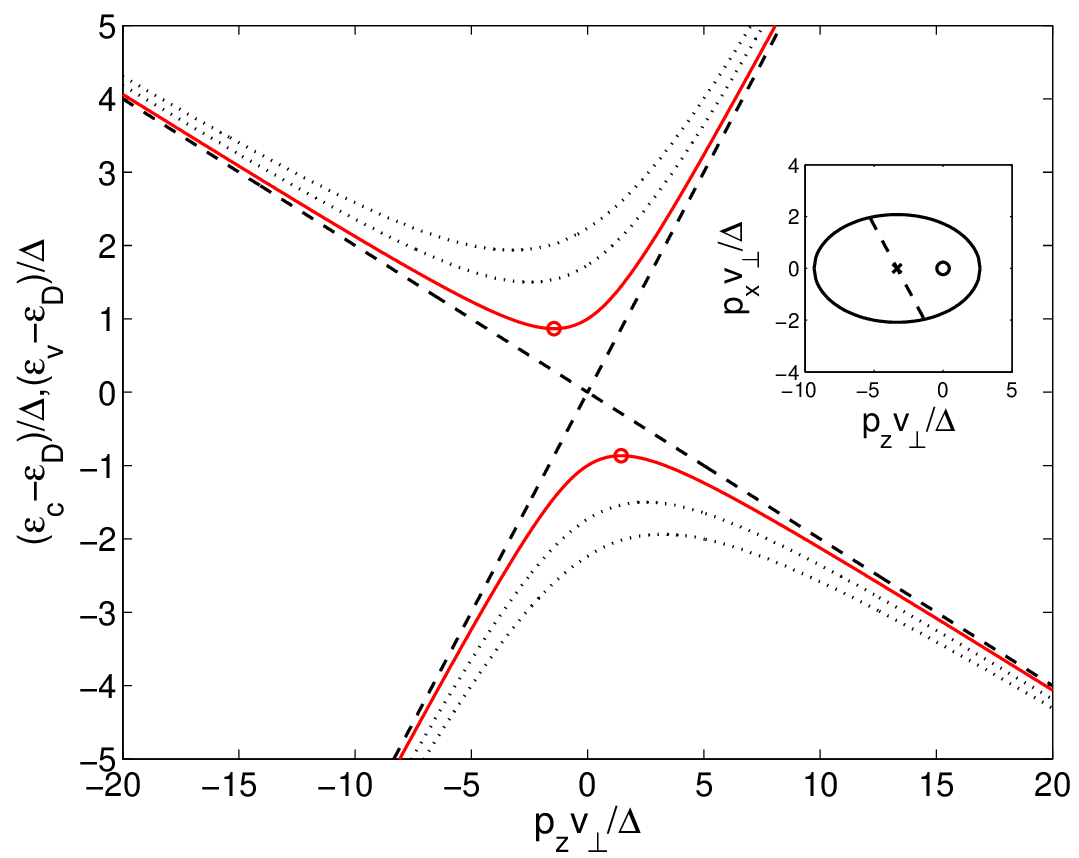}
\caption{\label{fig1} {\bf Dispersion of the two energy bands along the rotation axis in the vicinity of the quasi-Dirac point}. These bands $\varepsilon_{c}({\bf p})$ and $\varepsilon_{v}({\bf p})$ (red solid lines) are described by Eq.~(\ref{1}). The quasi-Dirac point (${\bf p}=0$) lies in the rotation axis $z$. The dashed lines corresponds to the case of the Dirac point ($\Delta=0$). The red circles mark the minimum of $\varepsilon_{c}({\bf p})$ and the maximum of $\varepsilon_{v}({\bf p})$. The minimal indirect gap $2\Delta_{\rm min}= 2\Delta(1-\tilde a_{z}^2)^{1/2}$ determined by these two points is less than $2\Delta$, the direct gap at ${\bf p}=0$. Here $v_x=v_y\equiv v_{\perp}$, $v_z/v_{\perp}=0.4$, $\tilde a_z=0.5$. The dotted lines show the Landau subbands $l=1, 2$ at $2e\hbar v_{\perp}^2H/ (c\Delta^2)=2$. Inset: The cross section (ellipse) of the Fermi surface by the plane $p_y=0$ at $(\varepsilon_{\rm F}-\varepsilon_{\rm D})/ \Delta=2$. The black dashed line marks the extremal cross section of the Fermi surface when the magnetic field lying in this plane is  directed at the angle $\theta=\pi/4$ to the $z$ axis. This cross section does not pass through the quasi-Dirac point ${\bf p}=0$ (black circle) at which the direct gap is minimal. The cross marks the center of the ellipse.
 } \end{figure}   

Formula (\ref{1}) gives the strict definition of the QD spectrum.
Now the question arises: What are conditions for existence of this spectrum in real crystals?
Note that Eq.~(\ref{1}) does not contain quadratic in $p_i$ terms in front of the square root and the terms $p_i^n$ with $n>2$ under the radical. This means that the effect of other bands on dispersion (\ref{1}) is negligible. This neglect is generally justified if (i) $\Delta$ is noticeably less than the energy spacing between $\varepsilon_{\rm D}$ and the remote bands at the point ${\bf p}=0$, and (ii) the strength of the spin-orbit interaction in the crystal, $\Delta_{\rm SO}$,  is significantly larger than  $\Delta$. The latter condition usually ensures non-small values of all three $v_i$ (at such $v_i$, the terms of higher orders in $p_i$ are relatively small). The strength of the SO coupling is also important for the spectrum of the charge carriers in the magnetic field.
(This spectrum depends both on dispersion (\ref{1}) and on the g factor of electron orbits; see below.) The value of $\Delta_{\rm SO}$ can be estimated from the band-structure calculations as a characteristic shift of the bands when the SO interaction is taken into account. If $\Delta$ itself has the spin-orbit origin, and $\Delta_{\rm SO}\sim \Delta$,  dispersion (\ref{1}) usually occurs only in the planes perpendicular to a certain line in the Brillouin zone (see Discussion).

\begin{figure}[tbp] 
 \centering  \vspace{+9 pt}
\includegraphics[scale=0.41]{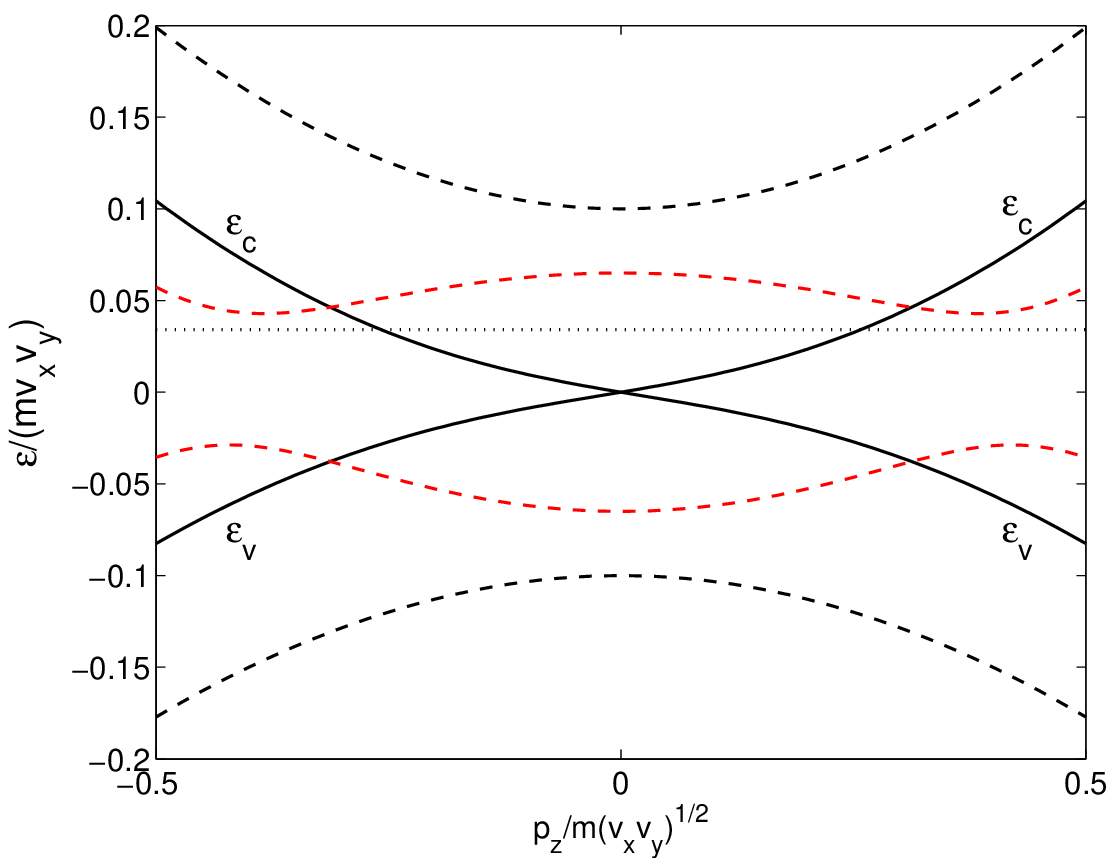}
\caption{\label{fig2} {\bf The two energy bands in the vicinity of  the middle of a $2$-fold rotation axis}. The rotation axis coincides with the $z$ axis (with $p_z=0$ being its middle). In the construction of the figure, dispersion relation (\ref{12}) with the first set of the parameters from Table \ref{tab2} has been used. The solid lines show the case of ZrTe$_5$ ($\Delta\approx 0$); the dotted line marks the position of the Fermi level in this material ($\varepsilon_{\rm F}\approx 36.1$ meV, $mv_xv_y\approx 1.06$ eV). Even below this level, a deviation of the dispersion from the linear law is visible. The black and red dashed lines correspond to the  bands when $\Delta/(mv_xv_y)=0.1$ and $\Delta/(mv_xv_y)=-0.065$ (i.e., before and after the band inversion), respectively. At $\Delta<\Delta_{\rm cr}=-v_z^2/(\alpha_c+\alpha_v)\approx -11.4$ meV, instead of the quasi-Dirac point with the gap $2|\Delta|$ at $p_z=0$, the two quasi-Dirac points with the gap $2[\Delta_{\rm cr}(2\Delta-\Delta_{\rm cr})]^{1/2}$ exist at $p_z^2=2(\Delta_{\rm cr}-\Delta)/(\alpha_c+\alpha_v)$.}
\end{figure}   

Consider now  how the quasi-Dirac spectra appear in crystals. As was mentioned in the Introduction, they can occur if a small gap develops in the true Dirac spectrum. However, like the Dirac points, pairs of the QD points can exist in $n$-fold rotation axes  when the control parameter $x$ lies in its certain interval (Fig.~\ref{fig2}). In a centrosymmetric crystal, all the electron states are doubly  degenerate in spin. In the axes, these degenerate states are  invariant under rotations through $2\pi/n$ angle. They are multiplied by factors $u_c$ and $u_c^*$ for the band $c$ and by  $u_v$ and $u_v^*$ for the band $v$ when such a rotation occurs \cite{yang}. Here $u_c$ and $u_v$ are complex numbers ($|u_c|=|u_v|=1$), and the asterisk marks the complex conjugate value. For the Dirac point to occur in the axis, the pairs $(u_c,u_c^*)$ and $(u_v,u_v^*)$ have to be different, otherwise, the states of the bands $c$ and $v$ are ``repulsed'' from each other \cite{yang}. Since for a $2$-fold rotation axis, the only possible  pair is $(i,-i)$, the Dirac points cannot appear in this case. The repulsion of the states can also occur in the $3$, $4$, and $6$-fold axes if the pairs coincide for the bands $c$ and $v$, i.e., the Dirac points cannot appear after the inversion of such bands. It turns out that in this case, the quasi-Dirac points arise; see Supplementary Note $1$. In other words, the inversion of the bands in any rotation axis yields either the Dirac or quasi-Dirac points, and we expect the QD points are fairly common in crystals. In particular, in the $2$-fold axis, only the pair of the quasi-Dirac points can occur.

{\bf The QD spectrum at nonzero magnetic fields.}
In the magnetic field $H$, the exact spectrum for the particles with dispersion relation (\ref{1}) is well known for the case of true relativistic electrons when ${\bf a}= 0$, $\Delta\neq 0$ \cite{LL4}. For the tilted dispersion when $0<\tilde a^2<1$, the spectrum was obtained at $\Delta=0$ \cite{m-sh,yu16,udag,tch} and  $\Delta\neq 0$ \cite{m-sh}. It is important that  this spectrum $\varepsilon^{(l)}(p_n)$ can be described with the equation \cite{m-sh},
 \begin{equation}\label{2}
S_{c,v}(\varepsilon^{(l)},p_n)=\frac{2\pi\hbar e H}{c}l,
\end{equation}
where $e$ is the absolute value of the electron charge;  $l=0,1,2,\dots$; $S_{c,v}(\varepsilon^{(l)},p_n)$ is the area of the cross section of the constant-energy surface $\varepsilon_{c,v}({\bf p})=\varepsilon^{(l)}$ by the plane $p_n=$ const.; the Landau levels (subbands) $\varepsilon^{(l)}(p_n)$ in Eq.~(\ref{2}) are double degenerate in spin for all $l\neq 0$, but the levels $l=0$ for the bands $c$ and $v$ are nondegenerate. Here $p_n$ is the component of the quasimomentum along the magnetic field. Note that in deriving formula (\ref{2}), the Zeeman term describing the direct interaction of the electron spin with the magnetic field was disregarded.

Formula (\ref{2}) looks like the semiclassical quantization condition, which, for a band doubly degenerate in spin, reads  \cite{Sh}:
 \begin{equation}\label{3}
S(\varepsilon^{(l)},p_n)=\frac{2\pi\hbar e H}{c}\left (l+\frac{1}{2}\pm \frac{gm_*}{4m} \right ),
\end{equation}
where $g$ is the g factor of a semiclassical orbit in the magnetic field, $m_*$ is the cyclotron mass of this orbit, and $m$ is the free-electron mass.
Equations (\ref{2}) and (\ref{3}) must coincide at least for large $l$ when the semiclassical approximation is valid. This means that $0.5\pm gm_*/4m$ are integer. A direct calculation \cite{g1,g2}  of the g factor for the dispersion law (\ref{1}) does confirm this conclusion. Moreover, since Eq.~(\ref{2}) reveals
the total coincidence of the semiclassical and exact spectra,
the nondegeneracy of the level $l=0$ and the double degeneracy of the Landau levels with $l\neq 0$ give  $|g|= |2m/m_*|$. Formula (\ref{2}) also means that the Landau subbands $\varepsilon^{(l)}(p_n)$ can be obtained from the semiclassical quantization condition for spinless particles \cite{Sh},
 \begin{equation} \label{4}
S_{c,v}(\varepsilon^{(l)},p_n)=\frac{2\pi\hbar e H}{c}\left (l+\gamma \right ),
\end{equation}
with constant $\gamma=0$ \cite{prl} different from the usual value $\gamma=1/2$ \cite{Sh}.

The coincidence of the semiclassical and exact spectra at all their quantum numbers is known for the two quantum systems \cite{LL3}.
They are  a harmonic oscillator (and hence, an electron with a parabolic dispersion) and an electron in the Coulomb field. Formula (\ref{2}) demonstrates that quasiparticles with dispersion (\ref{1}) give the third example of this total coincidence.

Consider the above result for the g factor in more detail.
This will help us evaluate the scope of applicability of formula (\ref{2}) to real situations. The area $S$ in quantization condition (\ref{3}) is defined by a dispersion relation, whereas the g factor of an electron orbit, e.g, in the band $c$, is determined by the following part $\hat H$ of the total electron Hamiltonian in the magnetic field ${\bf H}$ \cite{g1}:
 \begin{equation}  \label{5}
  \hat H= \frac{e}{2mc}{\bf H}(2{\bf L}_c^{\rm intra}+{\bf L}_c^{\rm c-v}+{\bf L}_c^{\rm c-rem}+2{\bf s}).
 \end{equation}
The contribution of the first term $2{\bf HL}_c^{\rm intra}$ to the g factor is expressed via the Berry phase $\Phi_{\rm B}$ of the orbit, $g^{\rm intra}=(2m/m_*)(\Phi_{\rm B}/\pi)$, where ${\bf L}_{c}^{\rm intra}$ is the intraband orbital electron moment. In the second term ${\bf HL}_c^{\rm c-v}$, the ${\bf L}_c^{\rm c-v}$ is the part of the orbital electron moment associated with virtual electron transitions between the bands $c$ and $v$. For ${\bf H}\parallel z$,
 \begin{equation*}
  {\bf L}_c^{\rm c-v}=\frac{\hbar m}{2i}\sum_{\rho''=1,2} \frac{(v_x)_{c\rho,v\rho''}(v_y)_{v\rho'',c\rho'} -(v_y)_{c\rho,v\rho''}(v_x)_{v\rho'',c\rho'}}{\varepsilon_v({\bf p})-\varepsilon_c({\bf p})},
 \end{equation*}
where $(v_i)_{c\rho,v\rho''}$ are the matrix elements of the velocity operator between the double degenerate states (marked by $\rho$, $\rho'$, $\rho''=1, 2$) of the bands $c$ and $v$. A similar expression corresponds to ${\bf L}_c^{\rm c-rem}$ in the third term ${\bf HL}_c^{\rm c-rem}$ which takes into account the virtual transitions between the $c$ and remote bands. The ${\bf L}_c^{\rm c-rem}$ is relatively small since it contains large denominators, $\varepsilon_{\rm rem}({\bf p})-\varepsilon_c({\bf p})$. This third term is usually of the order of the fourth term that describes the Zeeman interaction $e {\bf Hs}/mc$ of the electron spin ${\bf s}$ with the magnetic field. In the case of dispersion (\ref{1}), for which the remote bands are disregarded, the third term is absent, the fourth is neglected, and only the first two terms are actually taken into account in obtaining Eq.~(\ref{2}). It is necessary to emphasize that for various electron orbits in the vicinity of quasi-Dirac and Dirac points, the value of the Berry phase can be different. However the total contribution of the first two terms of the Hamiltonian  to the $g$ factor is always equal to $2m/m_*$ \cite{g2} [and hence, $\gamma=0$ in Eq.~(\ref{4})].

An estimate of the above contributions to the g factor shows that, like for Eq.~(\ref{1}), the necessary conditions for the applicability of formula (\ref{2}) to real situations are (i)  the relatively small $\Delta$, and (ii) the strong SO  interaction.
The importance of the strong SO interaction becomes clear from the following reasoning:
At $0<(\varepsilon_{\rm F}-\varepsilon_{\rm D})-\Delta_{\rm min}\ll \Delta$, Eq.~(\ref{1}) leads to a parabolic dependence  of $\varepsilon_c$ on ${\bf p}$ near the minimum of the $c$ band (Fig.~\ref{fig1}). This  dependence is typical for trivial (ordinary) charge carriers for which the g factor usually does not coincide with the specific value $2m/m_*$. As was emphasized above, the strength of the spin-orbit interaction, $\Delta_{\rm SO}$, is larger than $\Delta$ for the quasi-Dirac points. It is this condition that results in the specific value of the g factor. If $\Delta_{\rm SO}$ decreases and becomes less than $\Delta$, the value of the g factor begins to decrease, too. At small $\Delta_{\rm SO}/\Delta$, we obtain the estimate $g_{\rm orb}\sim (\Delta_{\rm SO}/\Delta)$ for the orbital part of the g factor  \cite{g1} that is determined by the first three terms in Hamiltonian (\ref{5}). Then, the total $g\sim 2+(\Delta_{\rm SO}/\Delta)$. Thus, we arrive at the case of the trivial charge carriers only for sufficiently small $\Delta_{\rm SO}$.

The above necessary conditions can be formulated as follows:
\begin{equation} \label{6}
\eta= \frac{\varepsilon_0}{E_0} \ll 1.
 \end{equation}
Here $\varepsilon_0 \sim {\rm max} (|\varepsilon_{\rm F}- \varepsilon_{\rm D}|,\Delta)$ is the characteristic scale of the energy spectrum under study, the scale $E_0$ is of the order of the  smallest value of the two energies: $mv_{\perp}^2$ and a gap between $\varepsilon_{\rm D}$ and the closest remote band at the point ${\bf p}=0$, $v_{\perp}$ is the typical charge-carrier velocity in the plane perpendicular to the magnetic field; see, e.g, Supplementary Eq.~(33).
If the parameter $\eta$ is not too small, additional terms that take into account the remote bands should be introduced into dispersion relation (\ref{1}). Beside this modification of Eq.~(\ref{1}), the last two terms in Eq.~(\ref{5}) produce a correction $\Delta g$ to the above universal value of the $g$ factor in semiclassical condition (\ref{3}), $g=2(m/m_*)+\Delta g$. This correction adds $\pm\Delta gm_*/4m$ to $\gamma=0$ in formula (\ref{4}) and leads to a splitting of the Landau levels with $l\ge 1$. The splitting $\Delta gm_*/2m$  is proportional to  the parameter $\eta$ since $|m_*|/m\approx |\varepsilon_{\rm F}- \varepsilon_{\rm D}|/ (mv_{\perp}^2)\sim \eta$. Strictly speaking, this estimate of the splitting is valid for low magnetic fields when many Landau levels lie under the Fermi surface, and the semiclassical approximation is accurate. However, the splitting is usually observed only for the lowest Landau levels  \cite{nara15,cao15,xiang15}, i.e., for strong $H$.
In this case, the correction to the g factor can increase even more  since  the lowest Landau levels for the modified dispersion (\ref{1}) are no longer described by the semiclassical formula.

To distinguish the Dirac electrons from the trivial charge carriers in crystals, a number of the physical effects are commonly used (see Introduction). Consider now these effects in the case of the quasi-Dirac spectra.

{\bf Quantum-oscillation phenomena.}
The quasi-Dirac spectra can be analyzed with the quantum-oscillation phenomena, e.g., with the de Haas-van Alphen and Shubnikov-de Haas effects.
In Supplementary Note $2$, formulas for the quantum-oscillation frequencies $F_i$ and the cyclotron masses $m_{*,i}$ are presented in the case of charge carriers with dispersion (\ref{1}). The subscript $i=x,y,z$ means that the appropriate quantity corresponds to the magnetic field directed along the $i$th axis.
For simplicity, consider the case  ${\bf a}= 0$.
Then, Supplementary  formulas (26)--(28) give
  \begin{eqnarray}\label{7}
 \frac{2e\hbar F_i}{c|m_{*,i}|}&=&\frac{(\varepsilon_{\rm F} - \varepsilon_{\rm D})^2-\Delta^2}{|\varepsilon_{\rm F} - \varepsilon_{\rm D}|}\equiv \tilde\varepsilon_{\rm F}, \\ \frac{F_i}{F_j}&=&\frac{m_{*,i}}{m_{*,j}}\equiv \epsilon_{ij},
 \label{8}
 \end{eqnarray}
where $\epsilon_{ij}=v_i/v_j$.
Besides, when the magnetic field lies in the $i-j$ plane at the angle $\theta$ to the $i$ axis, the $\theta$ dependence of the frequency $F$ has the form:
\begin{eqnarray}\label{9}
F(\theta)=\frac{F_i}{(\cos^2\theta + \epsilon_{ij}^2 \sin^2\theta)^{1/2}}.
\end{eqnarray}
As in the case of Dirac points \cite{m-sh19,m-sh21a}, relationship (\ref{7}) shows that the ratio $F_i/|m_{*,i}|$ is the same for the three directions of the magnetic field (in fact, it is the same for all directions of $H$). This key statement is true in the general case ${\bf a}\neq 0$ if $\Delta$ is replaced by $\Delta_{\rm min}$ (Supplementary Note 2).
Note that for trivial electrons with a parabolic dispersion, the ratio $F_i/|m_{*,i}|$ is also independent of the direction of the magnetic field. In this case, Eqs.~(\ref{8}), (\ref{9}) remain valid with $\epsilon_{ij}=(m_j/m_i)^{1/2}$ where $m_i$ are the effective masses of the parabolic spectrum. However, if $\varepsilon_{\rm F}$ and $F_i$ for such electrons change, e.g., due to the doping of the sample by impurities, their cyclotron masses remain unchanged. On the other hand, for the Dirac and quasi-Dirac points, one has  $m_{*,i}\propto \varepsilon_{\rm F}-\varepsilon_{\rm D}$. (For QD points, at $|\varepsilon_{\rm F}-\varepsilon_{\rm D}|\to \Delta_{\rm min}$, the decreasing $|m_{*,i}|$ reaches a constant proportional to $\Delta$.) Interestingly, without any doping of the sample, the $\varepsilon_{\rm F}$ dependence of the cyclotron masses can be revealed  by measuring the temperature correction to the $F_i$ \cite{guo,alexandr}.

The constant $\gamma$  (the g factor) in the semiclassical quantization condition determines the phase of the quantum oscillations and can be found in experiments \cite{Sh}. For example, the first harmonic of the magnetization produced by the quasi-Dirac point is proportional to $\sum_{\pm}\sin[2\pi (F_i/H- \phi_{\pm}) -\pi/4]$ where $\phi_{\pm}$ are the phases specified by Eq.~(\ref{3}),
 \[
 \phi_{\pm}=\frac{1}{2} \pm \frac{gm_*}{4m}.
 \]
Since $|gm_*|/4m=1/2$ for the QD points, the phases $\phi_{\pm}$ coincide with $\gamma$ introduced in Eq.~(\ref{4}), $\phi_{\pm}=\gamma=0$ (the phase values $0$ and $1$ are equivalent). We emphasize that according to Eq.~(\ref{2}), the above specific value of  $\gamma$ (of the g factor) is valid for any cross section of the Fermi surface surrounding the quasi-Dirac point. In particular, if the maximal cross section of the Fermi surface does not pass through the Dirac point ($H$ is not perpendicular to ${\bf a}\neq 0$, Fig.~\ref{fig1}), or if there is a nonzero gap $2\Delta$ in the spectrum, the Berry phase of the orbit differs from $\pi$. However, quantum-oscillations experiments should give $\gamma=0$  in both these cases. In other words, the phase of the quantum oscillation,  $\gamma$, rather than its constituent, the Berry phase, is robust with respect to small crystal-potential perturbations generating the gap.  Therefore, the measurements of the phase of the oscillations is the most direct way to distinguish the charge carriers near the QD points from the carriers with a dispersion different from Eq.~(\ref{1}).
However, these measurements do not differentiate between the Dirac and quasi-Dirac points.

Although the value of $\gamma$ is the hallmark of the quasi-Dirac points, in a number of the experiments with Cd$_3$As$_2$ \cite{cao15,xiang15} and ZrTe$_5$ \cite{yuan,liu2,wang18}, the measured $\gamma$ deviates from the zero value. The ratio $F_i/m_{*,i}$ for these materials also slightly depends on the direction of $H$ \cite{xiang15,yuan}, and hence $F_i/F_j$ does not coincide with $m_{*,i}/m_{*,j}$.
All these discrepancies between the theoretical and experimental results indicate that the dispersion of the charge carriers, at least  along one of the axes, deviates from Eq.~(\ref{1}), i.e., the parameter $\eta$ in  Eq.~({\ref{6}) is not small enough. In this case, a modification of the dispersion is required to describe experimental data. Below we will discuss this issue in more detail, using  ZrTe$_5$ as an example.

{\bf Thermodynamic quantities dependent on magnetic field.}
As an example of such thermodynamic quantities, consider magnetic susceptibility $\chi$, and compare $\chi$ produced by the Dirac and quasi-Dirac points. The magnetic susceptibility of the QD points was  calculated for weak  \cite{buot,m-sv,kosh15} and strong \cite{m-sh} magnetic fields. The specific case of the Dirac point was analyzed in a number of papers \cite{m-sv,m-sh,kosh15,m-sh16,m-sh19}.

The main results can be summarized as follows: The magnetic susceptibility for the Dirac point is diamagnetic and diverges logarithmically when the Fermi level $\varepsilon_{\rm F}$ approaches the Dirac energy, $\chi\propto \ln|\varepsilon_{\rm F}-\varepsilon_{\rm D}|$. This divergence is cut off at $|\varepsilon_{\rm F}-\varepsilon_{\rm D}| \sim {\rm max}(T,\Delta\varepsilon_H)$ where $\Delta\varepsilon_H=e\hbar H/(m_*c)$ is the spacing between the Landau levels, and $T$ is the temperature. For the case of the quasi-Dirac point, $\chi$ is the same, but this cut-off occurs at $|\varepsilon_{\rm F}-\varepsilon_{\rm D}| \sim {\rm max}(T,\Delta\varepsilon_H,\Delta_{\rm min})$ where $2\Delta_{\rm min}$ is the minimal indirect gap for dispersion (\ref{1}), Fig.~\ref{fig1}. Thus, the difference between the Dirac and the quasi-Dirac cases can manifest itself only at $T<\Delta_{\rm min}$ in the sufficiently low magnetic fields $\Delta\varepsilon_H<\Delta_{\rm min}$ and for the Fermi level lying in the gap,  $|\varepsilon_{\rm F}-\varepsilon_{\rm D}| < \Delta_{\rm min}$, or near it.

The similarity of the thermodynamic quantities for the Dirac and quasi-Dirac points can be understood from the following considerations: The characteristic feature of the Dirac spectrum is that the lowest Landau subband ($l=0$) is independent of $H$ and coincides with the dispersion law of the charge carriers along the direction of the magnetic field \cite{Moll}. It is clear from Eq.~(\ref{2}) that if there is a nonzero gap in the spectrum, the lowest Landau subband ($l=0$) still is independent of $H$.
That is why the thermodynamic quantities practically do not ``feel'' the gap. In contrast, for the trivial electrons, one has $\gamma=1/2$,   $S_{c,v}(\varepsilon^{(0)},p_n)= \pi\hbar e H/c$, and the lowest Landau subband $\varepsilon^{(0)}(p_n)$ depends on the magnetic field.

{\bf Fermi arcs, chiral anomaly, and magneto-optical conductivity.}
In a crystal with two Dirac points lying in a rotation axis, the  Fermi arcs can be observed on its surface \cite{xu}. The existence of these arcs is due to the nonzero topological invariant $\nu_{2D}$ defined in the plane which is perpendicular to the axis and passes through its middle $\Gamma$ \cite{yang,kargar}. This invariant is determined by the parities of the electron bands at TRIM in this plane of a crystal. When the inversion of the bands occurs at $\Gamma$, the invariant changes. This change  does not depend on whether the Dirac or the quasi-Dirac points appear in the axis after the inversion. Therefore, the arcs are expected to be observed in both these cases.

It is clear that the chiral anomaly \cite{niel} cannot occur for the quasi-Dirac points with a nonzero gap. However, the negative longitudinal magnetoresistance, which accompanies this anomaly, can be observed for the quasi-Dirac spectra \cite{andreev18}.  Moreover, the negative magnetoresistance can  arise even  in crystals with trivial charge carriers if in magnetic fields, these carriers are  redistributed  between their  Fermi-surface  pockets with different mobilities \cite{cichorek}.

Measurements of the magneto-optical conductivity make it possible to find the gap in electron spectra \cite{chen,yuan,martino,jiang20,mohelsky23,jiang23}, and therefore,
to distinguish between the quasi-Dirac and Dirac points. (The appropriate formulas in the case of the  QD spectrum are presented in Supplementary Note $3$.) However, if $|\varepsilon_{\rm F}-\varepsilon_{\rm D}|\gg \Delta$, a reliable detection of a nonzero $\Delta$ is obviously not an easy task.

{\bf Spectrum of ZrTe$_5$.}
To illustrate the above results, consider ZrTe$_5$.
Crystals of this material have an orthorhombic layered structure, in which the layers stack along the $b$ axis. As a result, ZrTe$_5$ shows a strong  anisotropy. The Dirac point can occur in the center $\Gamma$ of the Brillouin zone \cite{weng,fan}, i.e., the case (i)  mentioned in Introduction is realized in ZrTe$_5$. In absence of the magnetic field, the Hamiltonian of the charge carriers near the $\Gamma$ point has the form \cite{chen}:
\begin{eqnarray}\label{10}
\hat H_{\Gamma}\!&=&\hat H_{\rm diag}
+\!v_xp_x\tau_x\sigma_z\!+\!v_yp_y\tau_y\sigma_0\!+\! v_zp_z\tau_x\sigma_x,~~~~~
\end{eqnarray}
where  $\hat H_{\rm diag}=\Delta\tau_z\sigma_0$; $\sigma_i$ and $\tau_i$ are the Pauli matrices, whereas $\tau_0$ and $\sigma_0$ are the identity matrices (the $\tau$ and $\sigma$ mark the band and spin indices, respectively); $v_x$, $v_y$, $v_z$ are real constants with the dimension of a velocity;  $2\Delta$ is the gap in the spectrum (to obtain the true Dirac point, a small deformation of ZrTe$_5$ is required \cite{liu2,fan}); the energy is measured from the middle of the gap (i.e., $\varepsilon_{\rm D}=0$ below). The $z$ axis coincides with the $b$ axis,  whereas $x$ and $y$ are along the $a$ and $c$ axes of the crystal, respectively. The diagonalization of Hamiltonian (\ref{10}) yields dispersion (\ref{1}) with ${\bf a}=0$.

\begin{table}
\caption{\label{tab1} {\bf The frequencies and the cyclotron masses of the quantum oscillations in ZrTe$_5$.} The frequencies $F_i$ and the cyclotron masses $m_{*,i}$ (measured in the units of the free-electron mass) correspond to the magnetic fields directed along the axes $i=x,y,z$ of the crystal. The values of $F_i$ are found from  Figs.~4e, 4f in the paper of Yuan et al. \cite{yuan}, and  Fig.~5b \footnote{According to Fig.~5b in the work \cite{yuan}, the accuracy of the determination of $m_{*,z}$ is about $15\%$, whereas a possible error for $m_{*,x}$ and $m_{*,y}$ is near $5\%$.} presented there gives $m_i$. For each pair $F_i$, $m_{*,i}$, the ratio $\tilde\varepsilon_{\rm F}$ is calculated with Eq.~(\ref{7}); $\epsilon_{yx}=F_y/F_x$ here. }
\begin{tabular}{|c c c|ccc|ccc|c|}
\hline
\hline \\[-2.5mm]
$F_z$&$m_{*,z}$&$\tilde\varepsilon_{\rm F}$&$F_x$&$m_{*,x}$& $\tilde\varepsilon_{\rm F}$&$F_y$&$m_{*,y}$&$\tilde\varepsilon_{\rm F}$& $\epsilon_{yx}$ \\
T & &meV &T & &meV &T & &meV&  \\
 \colrule
5.3 &0.034 &36.1 &55.2 &0.266 &47.9 &32.4 &0.16  &46.9&0.587 \\
\hline \hline
\end{tabular}
\end{table}

Let us check whether experimental results on the quantum oscillations in this material confirm the existence of the quasi-Dirac point.
The Shubnikov - de Haas oscillations in ZrTe$_5$ were investigated in a number of works (see, e.g., papers  \cite{liu2,yuan,gaikwad,zhu} and references therein). The most full data  were presented by Yuan et al. \cite{yuan}. In Table \ref{tab1}, the frequencies of the oscillations and the cyclotron masses are written for the magnetic fields directed along the principal axes of the crystal \cite{yuan}. In this Table, the ratios $\tilde\varepsilon_{\rm F}$ calculated with Eq.~(\ref{7}) are also given. Although this ratio for the quasi-Dirac point should be independent of the orientation of $H$ within the accuracy of the experiment, the discrepancy between the obtained values of $\tilde\varepsilon_{\rm F}$ is about $30\%$, which slightly exceeds the possible experimental error.

Yuan et al. \cite{yuan} also measured the angular dependences of the oscillation  frequency when the magnetic field was rotated in the $z$-$x$ and $z$-$y$ planes. These dependences are well approximated  with formula (\ref{9}) (Fig.~\ref{fig3}), but the obtained values of $\epsilon_{zy}=F_z/F_y$ and  $\epsilon_{zx}= F_z/F_x$ are about $25\%$ less than  $m_{*,z}/m_{*,y}$ and $m_{*,z}/m_{*,x}$, respectively. On the other hand, the parameter of the anisotropy in the $x$-$y$ plane,  $\epsilon_{yx}= v_y/v_x$, is practically the same when it is calculated as $F_y/F_x$ or  $m_{*,y}/m_{*,x}$.
Using Supplementary Eq.~(22) for $F_{z}$ at $\tilde a_z=0$, the  values of $\epsilon_{yx}$ and the Fermi energy  $|\varepsilon_{\rm F}| \approx \tilde\varepsilon_{\rm F}=36.1$ meV from Table \ref{tab1}, we obtain the  velocities  $v_x\approx 5.6\times 10^{5}$ ms$^{-1}$ and  $v_y\approx 3.3\times 10^{5}$ ms$^{-1}$. (We take  $\tilde\varepsilon_{\rm F}\approx |\varepsilon_{\rm F}|$ since $\Delta$ is small in ZrTe$_5$ \cite{chen,yuan,martino,jiang20,mohelsky23,jiang23}.) Then, with  $v_z/v_x=\epsilon_{zx}=0.096$ found in Fig.~\ref{fig3}, we arrive at the estimate $v_z\approx  5.4\times 10^{4}$ ms$^{-1}$. Note that $\sqrt{v_xv_y}\approx 4.3\times 10^{5}$ ms$^{-1}$ is close to the values $(4.8- 5)\times 10^{5}$ ms$^{-1}$ obtained from magneto-optical measurements \cite{chen,martino,jiang20,mohelsky23,jiang23} and to $4.1\times 10^{5}$ ms$^{-1}$ found in the recent Shubnikov-de Haas experiment \cite{zhu}.

\begin{figure}[tbp] 
 \centering  \vspace{+9 pt}
\includegraphics[scale=0.45]{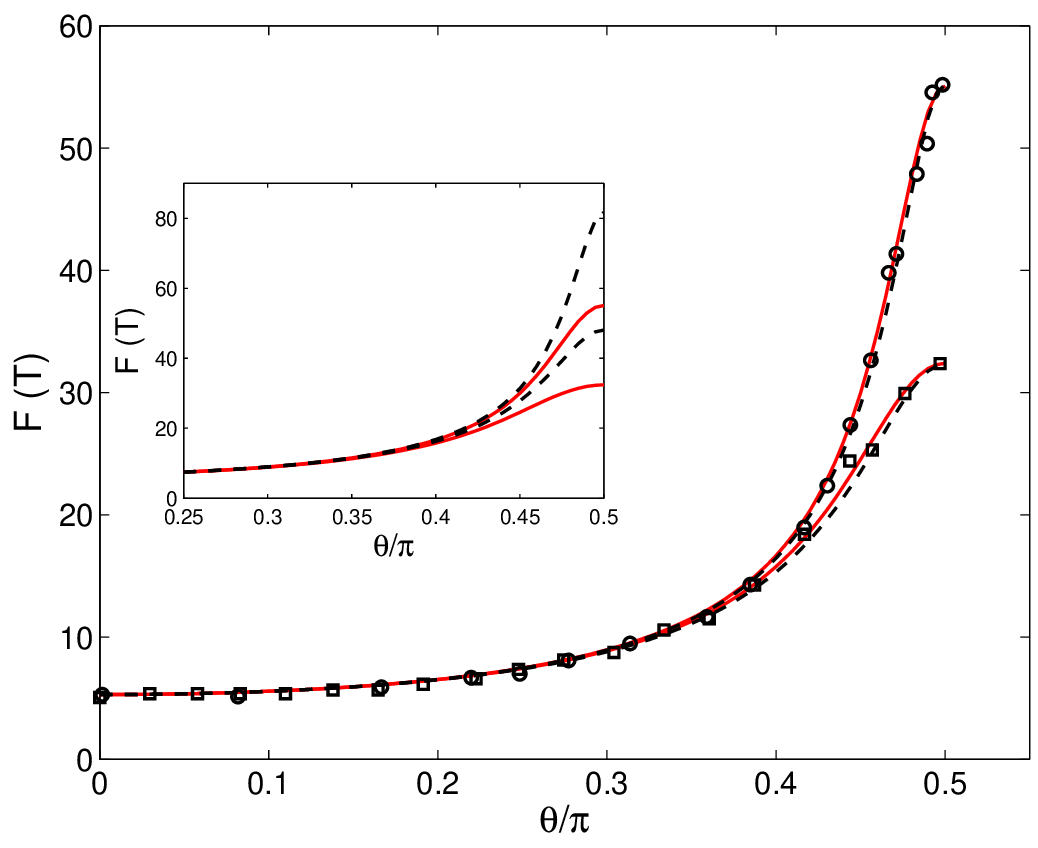}
\caption{\label{fig3} {\bf The angular dependences of the quantum-oscillations frequency for ZrTe$_5$.} The $\theta$ is  the angle   between the magnetic field $H$ and the $z$ axis. The experimental data \cite{yuan} for the frequency $F$ of the oscillations are shown for $H$ lying in the $z-x$  (cycles) and $z-y$ (squares) planes. For dispersion (\ref{1}), the dashed lines depict dependence (\ref{9}) with $\epsilon_{zx}=0.096$ and $\epsilon_{zy}=0.164$. The red solid lines correspond to dispersion (\ref{12}) and are calculated with Eqs.~(\ref{c1})--(\ref{c4}) and the first set of the parameters from Table \ref{tab2}. Inset: The red lines are the same as in the main panel. To clarify the role of the terms $\alpha_{i}p_z^2$ in dispersion (\ref{12}), the dashed lines are calculated with the same formulas as the red lines, but setting $\alpha_c=\alpha_v=0$ in the first set of the parameters [i.e., in fact, with Eq.~(\ref{9}) and $\epsilon_{zx}\approx 0.065$, $\epsilon_{zy}\approx 0.11$].
 } \end{figure}   

The above analysis of the frequencies and masses seems to indicate some  deviation of the electron spectrum along the $z$ axis in ZrTe$_5$ from the quasi-Dirac form. A more distinct support of this conclusion follows from  the measurements of the phase of the Shubnikov - de Haas oscillations \cite{liu2,yuan,wang18}. In particular, Yuan et al. \cite{yuan} found that at $H\parallel z$, this phase corresponds to the Dirac spectrum, whereas the phase changes by about $1/2$ when the direction of $H$ becomes almost perpendicular to the $z$ axis.

The deviation of the dispersion from Eq.~(\ref{1}) may be due to the layered structure of ZrTe$_5$. This layering leads to a relatively small velocity $v_z$ found above. To take into account the above-mentioned deviation, it is worth noting the following:
The symmetry of the point $\Gamma$ admits the terms of the form $\alpha_i p_i^2$ in the diagonal part of the Hamiltonian, $\hat H_{\rm diag}$, where $i=x, y, z$, and the constants $\alpha_i$ are of the order of $1/m$. However, the terms $\alpha_xp_x^2$, $\alpha_yp_y^2$ are relatively small as compared to $v_xp_x$, $v_yp_y$ and can be omitted. This is not the case for the quadratic term $\alpha_{z}p_z^2$  due to the relatively small $v_z$. Indeed, typical values of $p_z$ are determined by the relation $v_zp_z\sim \varepsilon_{\rm F}$, whereas the quadratic term at such $p_z$ is of the order of $(1/m)(\varepsilon_{\rm F}/v_z)^2\approx 2.2\varepsilon_{\rm F}$ for the values of $v_z$ and $\varepsilon_{\rm F}$  obtained above.  Therefore, the quadratic terms along the $z$ axis are important, and instead of $\hat H_{\rm diag}=\Delta\tau_z\sigma_0$, we will use the following expression:
\begin{eqnarray}\label{11}
\hat H_{\rm diag}=(\Delta+\frac{\alpha_{c}+\alpha_{v}}{2}p_z^2)\tau_z\sigma_0 +\frac{(\alpha_{c}-\alpha_{v})}{2}p_z^2\tau_0\sigma_0,
\end{eqnarray}
where $\alpha_{c}$ and $\alpha_{v}$ are some constants. Then, dispersion relation (\ref{1}) is replaced by
\begin{eqnarray}\label{12}
\varepsilon_{c,v}({\bf p})&=&\frac{\alpha_{c}-\alpha_{v}}{2}p_z^2 \nonumber \\
&\pm&\sqrt{(\Delta+\frac{\alpha_{c}+\alpha_{v}}{2}
p_z^2)^2 +v_{x}^2p_x^2 +v_{y}^2p_y^2+v_{z}^2p_z^2}.~~~~~
\end{eqnarray}
In fact, formula (\ref{12}) takes into account the effect of the remote bands on the dispersion of the $c$ and $v$ bands along the $z$ axis, and the quasi-Dirac spectrum occurs only in the $x-y$ plane. Curiously, if $\Delta$ is negative and $\Delta< \Delta_{\rm cr}= -v_z^2/(\alpha_c+\alpha_v)$, dispersion (\ref{12}) predicts the splitting of the quasi-Dirac point at $p_z=0$ into two QD points lying in the two-fold rotation axis $z$ (Fig.~\ref{fig2}).
It is necessary to note that the energy-band dispersion similar to Eq.~(\ref{12}) has already been proposed previously  \cite{martino,jiang20}. However, the term $v_z^2p_z^2$  was disregarded by Martino et al. \cite{martino}, and it was implied \cite{martino,jiang20} that $\alpha_{c}=\alpha_{v}$ although this restriction is not dictated by the symmetry of ZrTe$_5$.

\begin{table}
\caption{\label{tab2} {\bf The parameters of the spectrum for the charge carriers in ZrTe$_5$.}  The three sets correspond to different $\Delta$ in dispersion relation (\ref{12}). These $\Delta$ were found in experiments \cite{chen,yuan}, \cite{mohelsky23}, and \cite{jiang23}, respectively; $m$ is the free-electron mass.}
\begin{tabular}{c|c c c c c c c c|}
\hline
\hline \\[-2.5mm]
set&$\Delta$&$\varepsilon_{\rm F}$&$v_{x}$&$v_y$&$\alpha_c/\alpha_v$&$v_z$&$\alpha_cm$& $\alpha_vm$ \\
&meV&meV &$10^5$ ms$^{-1}$ &$10^5$ ms$^{-1}$& &$10^4$ ms$^{-1}$ & &   \\
 \colrule
$1$&$0$&$36.1$ &$5.61$ &$3.3$ &$1.3$&$3.64$ &$0.375$ &$0.289$  \\
$2$&$2.5$&$36.3$ &$5.61$ &$3.31$ &$1.35$&$3.36$&$0.363$ &$0.269$  \\
$3$&$-5$&$36.8$ &$5.66$ &$3.33$ &$1.3$&$4.38$ &$0.381$ &$0.293$  \\
\hline \hline
\end{tabular}
\end{table}

Interestingly, the Hamiltonian described by Eqs.~(\ref{10})--(\ref{12}) is equivalent to the Hamiltonian of McClure \cite{mcclure} for the electrons located near the point L of the Brillouin zone of Bi. In bismuth, the role of the $b$ axis plays the axis ``$2$'' directed along the length of the electron Fermi-surface pocket. This equivalence of the Hamiltonians means that results obtained with McClure model  for Bi can be extended to the case of ZrTe$_5$. In particular, experimental data on frequencies and cyclotron masses of pure bismuth and its alloys with Sb were successfully described  with  McClure model \cite{ponomarev}. We derive  similar formulas  in the case of dispersion relation (\ref{12}) and find the parameters of this relation (see Methods). In Table \ref{tab2}, the values of these parameters are presented for $\Delta \approx 0$ \cite{chen,yuan}, $\Delta=2.5$ meV \cite{mohelsky23} and $\Delta=-5$ meV \cite{jiang23} obtained in the magneto-optical experiments. Dependences $F(\theta)$ calculated with these formulas practically coincide with those given by Eq.~(\ref{9}), Fig.~\ref{fig3}. There is a tiny discrepancy between them only at $\cos\theta \sim v_z/v_x$.

\section{Discussion}

It was predicted \cite{weng,fan} that in ZrTe$_5$, the temperature expansion increases the  parameter $\Delta$, and the band inversion occurs at $\Delta=0$. This inversion corresponds to the transition from the strong topological insulator (STI), for which $\Delta<0$, to the weak topological insulator (WTI) characterized by a positive gap $2\Delta$. The existence of this transition and the ``initial'' state of ZrTe$_5$ at low temperatures are widely discussed in the literature; see recent papers \cite{mohelsky23,jiang23} and references therein. Unlike spectrum (\ref{1}), which does not depend on the sign of $\Delta$, dispersion (\ref{12}) makes it possible to distinguish between the STI and WTI phases, using the bulk properties of ZrTe$_5$. Indeed, as it was mentioned above, the critical value $2\Delta_{\rm cr}$ of the negative gap exists. At $\Delta<\Delta_{\rm cr}$, the quasi-Dirac point $\Gamma$ splits into the two QD points $\zeta$, which gradually shift along the $z$ axis with increasing $|\Delta|$ (Fig.~\ref{fig2}). Therefore, if two gaps is observed in magneto-optical measurements, this is indicative of the STI phase in ZrTe$_5$ \cite{jiang20}. The first gap is still equal to $2|\Delta|$, whereas the second gap $2|\Delta_{\zeta}|=2 \sqrt{\Delta_{\rm cr}(2\Delta-\Delta_{\rm cr})}$ corresponds to the two quasi-Dirac points. The observation of the two gaps at low temperatures was indeed reported for ZrTe$_5$ \cite{jiang20,jiang23}. In particular, Jiang et al.\ \cite{jiang20} obtained $2|\Delta|\approx 15$ meV and $2|\Delta_{\zeta}|\approx 11.2$ meV. These data lead to $|\Delta_{\rm cr}|\approx 2.5$ meV. At $v_z\approx 5\times 10^4$ ms$^{-1}$ \cite{jiang20} (which is comparable with $v_z$ from Table \ref{tab2}), the formula for $|\Delta_{\rm cr}|$ gives $m(\alpha_c+ \alpha_v) \approx 5.7$. This value is approximately an  order of magnitude  larger than those  following from Table \ref{tab2}. Although the doping of ZrTe$_5$  depends on the method of growing its single crystals \cite{martino}, the parameters $\alpha_c$ and $\alpha_v$ are determined by the remote bands. Hence, the large discrepancy between the values of $(\alpha_c+ \alpha_v)$ can hardly be explained by a difference in these methods for different experiments. To resolve this contradiction between the magneto-optical \cite{jiang20} and oscillation \cite{yuan} data, it would be useful to measure the three frequencies $F_i$ and the cyclotron masses $m_{*,i}$ of the Shubnikov-de Haas oscillations for the samples exhibiting the two gaps in magneto-optical experiments.

Keeping in mind the similarity of the electron Hamiltonians for Bi and ZrTe$_5$, let us point out a correspondence of certain results for these materials.
(i) As was mentioned above, the phase of the Shubnikov-de Haas oscillations sharply changes in ZrTe$_5$ when the magnetic-field direction approaches the $y$ axis \cite{liu2,yuan,wang18}. A similar change of the g factor for the electron orbits was observed in bismuth when $H$ was almost perpendicular  to the axis ``$2$'' \cite{Ed}. This angular dependence of the g factor in Bi was quantitatively described within McClure model \cite{g2}.
(ii) For ZrTe$_5$, the magnetic-field \cite{nair} and temperature \cite{singh} dependences of the magnetization were experimentally investigated at $H\parallel z$. For Bi, such dependences  were measured \cite{mcclure-sh,brandt77} and analyzed within McClure model \cite{m-sh00} many years ago. Thus, the approaches of the papers \cite{g2,m-sh00} can be useful to obtain additional information on the electron spectrum of ZrTe$_5$.

There is a similarity in the manifestations of the quasi-Dirac points in crystals with the strong spin-orbit interaction and of nodal lines in semimetals, for which this interaction is weak. Such lines mainly occur in neglect of the SO coupling \cite{armit,gao,m-sh19}. The coupling usually lifts the degeneracy of the bands, and the spectrum takes on the quasi-Dirac form (with small $\Delta=\Delta_{\rm SO}$) in any plane perpendicular to the line.
As in the case of the quasi-Dirac point, the g factor of an electron orbit surrounding the line in such a plane is the sum: $g=g_{\rm intra}+g_{\rm inter}$. Here $g_{\rm inter}$ and
$g_{\rm intra}= (2m/m_{*})(\Phi_{\rm B}/\pi)$  are associated with the interband electron orbital moment and with the Berry phase $\Phi_{\rm B}$ of the orbit, respectively. The total g factor again has the universal value $2m/m_*$, and $\gamma=0$ in quantization condition (\ref{4}) \cite{jetp}.
This result justifies the concept of the nodal lines since it ensures  stability of their physical properties with respect to the lifting of the degeneracy. Moreover, when the radius of the orbit (i.e., $|\varepsilon_{\rm F}-\varepsilon_{\rm D}|$) increases, one has $\Phi_{\rm B}\to \pi$ and $g_{\rm inter}/g \sim \Delta/ |\varepsilon_{\rm F} -\varepsilon_{\rm D}|\to 0$. Thus, already near the nodal line, the constant $\gamma=0$ can be represented as $1/2-\Phi_{\rm B}/(2\pi)$ with $\Phi_{\rm B}= \pi$.
It is important that in crystals with the weak SO interaction, these values of $\Phi_{\rm B}$ and $\gamma$ do not depend on the size and shape of the electron orbit  \cite{prl}. This result explains why the properties of nodal-line semimetals (e.g., the drumhead surface  states) remain valid far away from the line.
Note that for the quasi-Dirac  points in crystals with the strong SO interaction, the same  universal value $\gamma=0$ occurs only near these points where Eq.~(\ref{1}) accurately describes the electron spectrum. The same statement is true for the Weyl points \cite{m-sh22}.

\section{Conclusions}

The Landau levels for charge carriers located near the quasi-Dirac and Dirac points are very similar. This similarity leads to a practical coincidence of the physical phenomena determined by these points. In the case of ZrTe$_5$, the published experimental data on the Shubnikov-de Haas effect indicate that the spectrum of this material has the quasi-Dirac form only in the plane of its layers. In the direction perpendicular to them, the real charge-carrier dispersion deviates from formula (\ref{1}). Using ZrTe$_5$ as an example, we show how such a deviation can be taken into account to describe experimental data.

\section{Methods}

{\bf Determination of the parameters describing the charge carriers in ZrTe$_5$.}
The frequencies $F_x$, $F_y$ and the cyclotron masses $m_{*,x}$, $m_{*,y}$  can be calculated analytically  in the case of dispersion relation (\ref{12}). In particular, we obtain,
\begin{eqnarray}\label{c1}
\frac{2\pi e \hbar F_x}{c}&=&\frac{4(\varepsilon_{\rm F}^2- \Delta^2)^{3/4}(2R)^{3/2}}{3v_y(\alpha_c\alpha_v)^{1/4}}[(1\!-\!k^2)K(k)\! \nonumber \\
&+&\!(2k^2\!-\!1)E(k)], \\
m_{*,x}&=&\frac{2\varepsilon_{\rm F}}{\pi v_y(\alpha_c\alpha_v)^{1/4}(\varepsilon_{\rm F}^2- \Delta^2)^{1/4}(2R)^{1/2}}\Big[K(k)  \nonumber \\
&-&\frac{(\alpha_c-\alpha_v)(\varepsilon_{\rm F}^2- \Delta^2)^{1/2}R}{\varepsilon_{\rm F}(\alpha_c\alpha_v)^{1/2}} [K(k)\!-\!(1\!-\!k^2)E(k)]\Big],~~~~~ \label{c2}
\end{eqnarray}
where $K(k)$ and $E(k)$ are the complete elliptic integrals of the first and second kinds, respectively,
\begin{eqnarray}\label{c3}
k^2=\frac{0.5B+R}{2R},\ \ \  R=\sqrt{1+\frac{B^2}{4}},  \\
B=-\frac{\varepsilon_{\rm F}(\alpha_c-\alpha_v)+\Delta(\alpha_c+\alpha_v) +v_z^2}{\sqrt{\alpha_c\alpha_v(\varepsilon_{\rm F}^2- \Delta^2)}}. \nonumber
\end{eqnarray}
Formulas (\ref{c1})--(\ref{c3}) agree with the expressions derived in the case of bismuth \cite{mcclure}. Formulas for $F_y$ and $m_{*,y}$ are obtained by the replacement of $v_y$ by $v_x$ in Eqs.~(\ref{c1}) and (\ref{c2}). This replacement shows that the relation $F_x/F_y=m_{*,x}/m_{*,y}=v_x/v_y$ remains true in the case of dispersion (\ref{12}), cf. Eq.~(\ref{8}). As to $F_z$ and $m_{*z}$, they are the same for dispersion relations (\ref{1}) and (\ref{12}) and are described by Supplementary  formulas (22) and (23) with $\tilde a_z=0$. Therefore, using these two formulas and the equality  $F_x/F_y=v_x/v_y$, one can find $\varepsilon_{\rm F}$, $v_x$, $v_y$ if $\Delta$ is known. In Table \ref{tab2}, we present the values of these parameters for $\Delta\approx 0$ \cite{chen,yuan}, $\Delta=2.5$ meV \cite{mohelsky23}, and $\Delta=-5$ meV \cite{jiang23}.

\begin{figure}[tbp] 
 \centering  \vspace{+9 pt}
\includegraphics[scale=0.45]{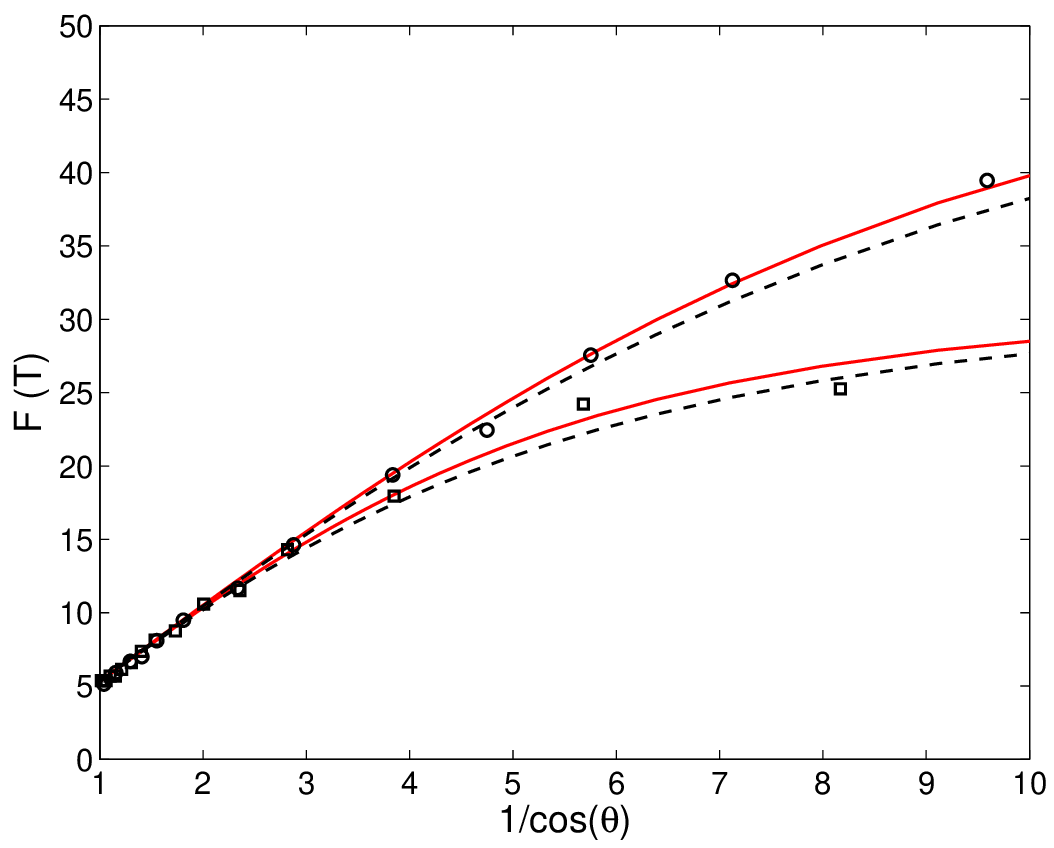}
\caption{\label{fig4} {\bf Another representation of the angular dependences shown in the main panel of Fig.~\ref{fig3}.} Here $F$ and $\theta$ are the frequency of the oscillations and the angle between the magnetic field $H$ and the $z$ axis, respectively. The experimental data \cite{yuan} are shown for $H$ lying in the $z-x$  (cycles) and  $z-y$ (squares) planes. The red  solid lines are calculated with Eqs.~(\ref{c1})--(\ref{c4}) and the first set of the parameters from Table \ref{tab2} (the second and third sets provide the same accuracy of the fit to the data). The dashed lines are described by Eq.~(\ref{9}) with $\epsilon_{zx}=0.096$,  $\epsilon_{zy}=0.164$.
 } \end{figure}   

If the magnetic field is at the angle $\theta$ to the $z$ axis, formulas for $F(\theta)$ and  $m_{*}(\theta)$ at $H$ lying in the $x-z$ and $y-z$ planes are obtained from Eqs.~(\ref{c1})--(\ref{c3}) by the following substitutions:
\begin{eqnarray}\label{c4}
\alpha_{c,v}\rightarrow \alpha_{c,v}\sin^2\!\theta,\ \ \ v_z^2 \rightarrow v_z^2\sin^2\!\theta+v_x^2\cos^2\!\theta,\\
\alpha_{c,v}\rightarrow \alpha_{c,v}\sin^2\!\theta,\ \ \
 v_z^2 \rightarrow v_z^2\sin^2\!\theta+v_y^2\cos^2\!\theta,
 \nonumber
 \end{eqnarray}
respectively. When $\alpha_c$ and $\alpha_v$ are relatively small, i. e., when the parameter $\eta=\sqrt{\alpha_c\alpha_v}\varepsilon_{\rm F}/v_z^2\ll 1$,  we obtain the following expressions from Eqs.~(\ref{c1})--(\ref{c4}):
 \begin{eqnarray*}
 B&\approx& -\frac{v_z^2+v_x^2\cot^2\!\theta}{\sqrt{\alpha_c\alpha_v (\varepsilon_{\rm F}^2-\Delta^2)}},\ \ \ |B|\gg 1,\ \ \
 2R\approx\!|B|+\frac{2}{|B|},\\
  k^2&\approx&\!\frac{1}{B^2}\ll 1,\ \ \ K(k)\!\approx\! \frac{\pi}{2}(1\!+\!\frac{k^2}{4}),\ \ \ E(k)\!\approx\! \frac{\pi}{2}(1\!-\!\frac{k^2}{4}), \\
  \frac{2\pi e \hbar F(\theta)}{c}\!&\approx&\!\frac{4(\varepsilon_{\rm F}^2- \Delta^2)^{3/4}|B|^{3/2}}{3v_y(\alpha_c\alpha_v)^{1/4}\sin\theta} \frac{3\pi}{4B^2}\!\approx\!\frac{\pi(\varepsilon_{\rm F}^2- \Delta^2)}{v_y\!\sqrt{v_z^2\sin^2\!\theta\!+\!v_x^2\cos^2\!\theta}}, \\
  m_{*}(\theta)\!&\approx&\!\frac{\varepsilon_{\rm F}}{v_y \sqrt{v_z^2\sin^2\!\theta\!+\!v_x^2\cos^2\!\theta}}.
   \end{eqnarray*}
The last two formulas reduce to Supplementary Eqs.~(22), (23) at $\theta=0$ and to  Eq.~(\ref{9}) at $\theta\neq 0$.

If $\alpha_c$ and $\alpha_v$ increase,  the parameter $\eta$ can become of the order of unity. This means that the dispersion of the bands along the $z$ axis noticeably deviates from the quasi-Dirac form. Nevertheless, we still have $F(\theta)\approx F(0)/\cos\theta$ at the angles $\theta\lesssim 1$ due to the large value of $v_x/v_z$. However, when $\theta$ is close to $\pi/2$ (when $\cot\theta\sim v_z/v_x$), the dependence $F(\theta)$ in this region of the angles becomes sensitive to the value of $\alpha_c/\alpha_v$. If $\alpha_c/\alpha_v$ is large, the parameter $k^2$ remains small at $\theta\to \pi/2$ [see formula for the parameter $B$ in Eqs.~(\ref{c3})], and $F(\theta)$ tends to the universal form given by Eq.~(\ref{9}) (the dashed lines in Fig.~\ref{fig4}). If $\alpha_c/\alpha_v$ decreases, $F(\theta)$ deviates from dependence (\ref{9}).  Therefore, a precise  measurement of $F(\theta)$ in the region  $\cos\theta\lesssim v_z/v_x$ enables one to find the value of $\alpha_c/\alpha_v$.

For ZrTe$_5$, the dependences $F(\theta)$ were measured for the magnetic field lying in the $z-x$ and $z-y$ planes \cite{yuan}. These dependences are really close to $F(0)/\cos\theta$ at $(1/\cos\theta) \lesssim 3$, but they deviate from this simple dependence when $\theta$ tends to $\pi/2$ (Fig.~\ref{fig4}). There is also a slight deviation of the experimental data from the universal form given by Eq.~(\ref{9}).

In order to determine the values of $v_z$, $\alpha_c$, and $\alpha_v$ for the ZrTe$_5$, it is convenient to use formula (\ref{c1}) and an expression for $2\pi e\hbar F_x \varepsilon_{\rm F}/[cm_{*,x} (\varepsilon_{\rm F}^2- \Delta^2)]$ that is obtained as the ratio of Eqs.~(\ref{c1}) and (\ref{c2}). We also impose the requirement that the calculated dependences $F(\theta)$ provide the best fit to the experimental data \cite{yuan} in the interval $3<1/\cos\theta<10$.
With these three conditions, we find the values of the parameters $v_z$, $\alpha_c$, and $\alpha_v$ presented in Table \ref{tab2}.
Interestingly, for $F_z$, $m_{*,z}$, $F_x$, $m_{*,x}$, $\epsilon_{yx}$ from Table \ref{tab1}, the first two conditions can be satisfied only at $\alpha_c/\alpha_v\ge 0.81$, $0.95$, $1.02$, if $\Delta=-5$, $0$, $2.5$ meV, respectively.

\renewcommand{\tablename}{Supplementary Table}
\setcounter{table}{0}

\section{Supplementary information}

\subsection{Supplementary Note $1$: Dirac and quasi-Dirac points in  $n$-fold rotation axes}

Let us discuss how a pair of the quasi-Dirac points can appear in the $n$-fold rotation axis (denoted as the $z$ axis below) of the Brillouin zone for a crystal with  inversion symmetry and strong spin-orbit interaction. Due to the time reversal and inversion symmetries, all the electron states are doubly  degenerate in spin in such a crystal.  In the axis, these degenerate states are invariant under rotations through $2\pi k/n$ angles where $k$ is an integer. The states at the point $\Gamma$, which is the middle of the  axis (or the crossing point of the axis with the surface of the Brillouin zone), have an additional symmetry since this point is invariant under the inversion, and so the electron states of all the electron-energy bands have a certain parity $p=\pm 1$ at $\Gamma$. To understand the origin of the Dirac and quasi-Dirac points, we will consider possible ${\bf k}\cdot {\bf p}$ Hamiltonians for two close bands at the point $\Gamma$ and analyze the case of the inversion of the bands.

Taking into account that all states in the crystals are invariant relative to the transformation $U=I(\sigma_yK)$ where $\sigma_y$ is the Pauli matrix and $I$, $K$ are the operators of the inversion and the complex conjugation \cite{bir}, respectively, all such Hamiltonians have the form:
\begin{eqnarray}\label{1a}
\hat H=\left (\begin{array}{cccc} {\mathcal E}_{i} & 0 & t & u \\ 0 & {\mathcal E}_{i} & -u^* & t^* \\t^* & -u & {\mathcal E}_{j} & 0 \\ u^*& t & 0 & {\mathcal E}_{j} \\
\end{array} \right),
 \end{eqnarray}
where
\begin{eqnarray}\label{2a}
{\mathcal E}_{i}&=&\Delta+ \frac{\alpha_{i}^{z}}{2}p_z^2+ \frac{\alpha_{i}^x}{2}p_x^2 + \frac{\alpha_{i}^y}{2}p_y^2,\\
{\mathcal E}_{j}&=&-\Delta -\frac{\alpha_{j}^{z}}{2}p_z^2- \frac{\alpha_{j}^x}{2}p_x^2 - \frac{\alpha_{j}^y}{2}p_y^2, \nonumber
\end{eqnarray}
$\alpha_{i}^{z}$, $\alpha_{j}^{z}$, $\alpha_{i}^{x}$, $\alpha_{j}^{x}$, $\alpha_{i}^{y}$, $\alpha_{j}^{y}$ are some real constants, $2\Delta$ is gap between the bands $i$ and $j$ at the point $\Gamma$, and $t$ and $u$ are the quasimomentum combinations that are determined by the point group of $\Gamma$.
A diagonalization of Hamiltonian (\ref{1a}) - (\ref{2a})  yields the dispersion relation for the bands $\varepsilon_{i}({\bf p})$ and $\varepsilon_{j}({\bf p})$,
 \begin{eqnarray}\label{3a}
& &\varepsilon_{i,j}({\bf p})=\frac{\alpha_{i}^z-\alpha_{j}^z}{2}p_z^2 +\!\frac{\alpha_{i}^{x}-\alpha_{j}^{x}}{2}p_x^2+ \frac{\alpha_{i}^{y}-\alpha_{j}^{y}}{2}p_y^2 \nonumber  \\
&\pm&\Big[\big[\Delta+\!\frac{\alpha_{i}^z+\alpha_{j}^z}{2}p_z^2 +\!\frac{\alpha_{i}^{x}+\alpha_{j}^{x}}{2}p_x^2 +\!\frac{\alpha_{i}^{y}+\alpha_{j}^{y}}{2}p_y^2\big]^2 \nonumber \\
&+& |t|^2\!\!+\!|u|^2\Big]^{1/2}.
\end{eqnarray}
The inversion of the bands in the $p_z$ axis occurs when $\Delta(\alpha_{i}^z+\alpha_{j}^z)<0$. Without loss generality, we will assume that $\alpha_{i}^z+\alpha_{j}^z$ is always positive and  $\Delta<0$ for the inverted bands. As an example, consider the case of the $4$-fold axis. (In this case, one has $\alpha_{i,j}^x=\alpha_{i,j}^y\equiv \alpha_{i,j}^{\perp}$). All the other rotation axes can be analyzed in similar manner. For the $4$-fold axis, the point group of
$\Gamma$ can be ${\bf C}_{4h}$ or ${\bf D}_{4h}$. (We use the same notations of the groups as in Refs.~\cite{LL3}).

\begin{table}\label{tab1a}
\caption{\textbf{The spinor representations of the point group ${\bf C}_{4h}$}. Here $E$, $I$, $C_n$ are the identity transformation, the inversion, and the rotation through the $\pi/2$ angle, respectively; $\epsilon=exp(i\pi/4)$ and $\epsilon^*=exp(-i\pi/4)=-exp(3i\pi/4)$.}
\begin{tabular}{c|cccccccc}
\hline
\hline \\[-2.5mm]
 &$E$&$C_4$&$C_2$&$C_4^3$&$I$&$IC_4$ &$IC_2$&$IC_4^3$ \\
 \colrule
$A'_+$&$1$&$\epsilon$&$i$&$-\epsilon^*$&$1$&$\epsilon$&$i$&$-\epsilon^*$ \\
$A'_-$&$1$&$\epsilon$&$i$&$-\epsilon^*$&$-1$&$-\epsilon$&$-i$&$\epsilon^*$ \\
$B'_{1+}$&$1$&$-\epsilon$&$i$&$\epsilon^*$&$1$&$-\epsilon$&$i$&$\epsilon^*$ \\
$B'_{1-}$&$1$&$-\epsilon$&$i$&$\epsilon^*$&$-1$&$\epsilon$&$-i$&$-\epsilon^*$ \\
$B'_{2+}$&$1$&$-\epsilon^*$&$-i$&$\epsilon$&$1$&$-\epsilon^*$&$-i$&$\epsilon$ \\
$B'_{2-}$&$1$&$-\epsilon^*$&$-i$&$\epsilon$&$-1$&$\epsilon^*$&$i$&$-\epsilon$ \\
$B'_{3+}$&$1$&$\epsilon^*$&$-i$&$-\epsilon$&$1$&$\epsilon^*$&$-i$&$-\epsilon$ \\
$B'_{3-}$&$1$&$\epsilon^*$&$-i$&$-\epsilon$&$-1$&$-\epsilon^*$&$i$&$\epsilon$ \\
\hline \hline
\end{tabular}
\end{table}

\begin{table}\label{tab2a}
\caption{\textbf{Multiplication table for the spinor representations of the point group ${\bf C}_{4h}$.} The results of the multiplications are the ordinary representations indicated in Supplementary Table III. The last column gives the symbolic designations of the doubly degenerate bands and the appropriate factors $u_l$, $u_l^*$; the signs $\pm$ in the subscripts indicate the parity of these bands.}
\begin{tabular}{c|cc|cc|cc|cc|c}
\hline
\hline \\[-2.5mm]
 &$A'_+$&$B'_{3+}$&$A'_-$&$B'_{3-}$&$B'_{1+}$&$B'_{2+}$ &$B'_{1-}$&$B'_{2-}$& band \\
 \colrule
$(A'_+)^*$&$A_+$&$B_{3+}$&$A_-$&$B_{3-}$&$B_{1+}$&$B_{2+}$&$B_{1-}$&$B_{2-}$& $AB_+$ \\
$(B'_{3+})^*$&$B_{2+}$&$A_+$&$B_{2-}$&$A_-$&$B_{3+}$&$B_{1+}$&$B_{3-}$&$B_{1-}$& $(\epsilon,\epsilon^*)$\\
\colrule
$(A'_-)^*$&$A_-$&$B_{3-}$&$A_+$&$B_{3+}$&$B_{1-}$&$B_{2-}$&$B_{1+}$&$B_{2+}$& $AB_-$ \\
$(B'_{3-})^*$&$B_{2-}$&$A_-$&$B_{2+}$&$A_+$&$B_{3-}$&$B_{1-}$&$B_{3+}$&$B_{1+}$& $(\epsilon,\epsilon^*)$\\
\colrule
$(B'_{1+})^*$&$B_{1+}$&$B_{2+}$&$B_{1-}$&$B_{2-}$&$A_+$&$B_{3+}$&$A_-$&$B_{3-}$& $B_{12+}$ \\
$(B'_{2+})^*$&$B_{3+}$&$B_{1+}$&$B_{3-}$&$B_{1-}$&$B_{2+}$&$A_+$&$B_{2-}$&$A_-$& $(-\epsilon,-\epsilon^*)$\\
\colrule
$(B'_{1-})^*$&$B_{1-}$&$B_{2-}$&$B_{1+}$&$B_{2+}$&$A_-$&$B_{3-}$&$A_+$&$B_{3+}$& $B_{12-}$ \\
$(B'_{2-})^*$&$B_{3-}$&$B_{1-}$&$B_{3+}$&$B_{1+}$&$B_{2-}$&$A_-$&$B_{2+}$&$A_+$& $(-\epsilon,-\epsilon^*)$\\
\hline \hline
\end{tabular}
\end{table}

\subsubsection{Group ${\bf C}_{4h}$}

Let us begin with the case when the group of $\Gamma$ is ${\bf C}_{4h}$. All the spinor representations of this group are one-dimensional \cite{bir}, see Supplementary Table I. Due to the time reversal symmetry, the spinor representations with complex conjugate characters must be combined into pairs that just provide the representations of doubly degenerate  bands; see Supplementary  Table II. When the rotation through $\pi/2$ angle occurs, the two degenerate states of the band $l$ ($l=i$ or $j$) are multiplied by the factors $u_l$ and $u_l^*$. Here $u_l$ is a complex number  with $|u_l|=1$, and the asterisk marks the complex conjugate value. These ($u_l$,$u_l^*$) are indicated in the last column of Supplementary Table II. Note that the bands with the coinciding $(u_l,u_l^*)$ have opposite parities $p$ at the point $\Gamma$. Therefore, such bands are distinguishable with their values of $p$, and we may consider the inversion of the bands even if their $(u_l,u_l^*)$ are the same.

The direct products of the spinor representations give the ordinary representations of the group, see Supplementary Tables II and III. These ordinary representations determine nonzero combinations of the quasimomentum in the appropriate matrix elements of Hamiltonian (\ref{1a}) \cite{LL3,bir}. These combinations are also indicated in Supplementary Table III. For the diagonal cells of Supplementary Table II, only the identity representation $A_+$ gives the nonzero contributions to ${\mathcal E}_{i}$ and ${\mathcal E}_{j}$ \cite{LL3}, and these contributions are explicitly taken into account in formulas (\ref{2a}).

Let the bands $AB_+$ and $B_{12-}$ be close to each other at the point $\Gamma$. Then, Supplementary Tables II and III yield: $t=c_tp_z(p_x^2-p_y^2)$, $u=v_{\perp}(p_x-ip_y)$, and \begin{eqnarray}\label{4a}
|t|^2+|u|^2=c_t^2p_z^2(p_x^2-p_y^2)^2+v_{\perp}^2(p_x^2+p_y^2),
\end{eqnarray}
where $c_t$ and $v_{\perp}$ are some real constants. When the inversion of bands occurs in the $p_z$ axis ($p_x=p_y=0$), the pair of the Dirac points appear at $p_z=\pm p_{z0}$,
 \begin{eqnarray}\label{5a}
 p_{z0}=\sqrt{-\frac{2\Delta}{(\alpha_{i}^z+\alpha_{j}^z)}},
\end{eqnarray}
where this $p_{z0}$ is determined by the condition ${\mathcal E}_{i}(p_{z0})-{\mathcal E}_{j}(p_{z0})=0$.
In the vicinity of $p_{z0}$, the dispersion defined by Eqs.~(\ref{3a}), (\ref{4a}) reduces to the form:
 \begin{eqnarray}\label{6a}
\varepsilon_{i,j}({\bf p})&=&\varepsilon_d +a_z (p_z-p_{z0}) \nonumber \\
&\pm&\sqrt{\tilde v_z^2(p_z-p_{z0})^2
+v_{\perp}^2(p_x^2+p_y^2)},
\end{eqnarray}
where we have omitted the terms of  higher orders in $p_x$, $p_y$,  $p_z-p_{z0}$, and
 \begin{eqnarray}\label{7a}
a_z&=&(\alpha_{i}^z-\alpha_{j}^z)p_{z0}, \ \ \  \varepsilon_d=\frac{(\alpha_{i}^z-\alpha_{j}^z)}{2}p_{z0}^2,\\
\tilde v_z^2&=&(\alpha_{i}^z+\alpha_{j}^z)^2p_{z0}^2. \label{8a}
  \end{eqnarray}
The same Dirac points appear at the inversion of the bands $AB_-$ and $B_{12+}$. In the case when the band $i$ and $j$ coincide with $AB_+$ and $B_{12+}$ or with $AB_-$ and $B_{12_-}$, Supplementary Tables II and III give,
 \begin{eqnarray}\label{9a}
|t|^2+|u|^2=c_t^2(p_x^2-p_y^2)^2+c_u^2p_z^2(p_x^2+p_y^2),
\end{eqnarray}
where $c_t$, $c_u$ are some real constants. The pair of the Dirac points again appears at $p_z=\pm p_{z0}$ determined by Eq.~(\ref{5a}), and in the vicinity of the $p_{z0}$, the dispersion defined by Eqs.~(\ref{3a}), (\ref{9a}) reduces to the form:
 \begin{eqnarray}\label{10a}
\varepsilon_{i,j}({\bf p})&=&\varepsilon_d +a_z (p_z-p_{z0}) \nonumber \\
&\pm&\sqrt{\tilde v_z^2(p_z-p_{z0})^2
+\tilde v_{\perp}^2(p_x^2+p_y^2)},
\end{eqnarray}
where $\varepsilon_d$, $a_z$, $\tilde v_z^2$ are still described by Eqs.~(\ref{7a}), (\ref{8a}), and
 \begin{eqnarray}\label{11a}
\tilde v_{\perp}^2=c_u^2p_{z0}^2.
  \end{eqnarray}
Note that in the cases of Eqs.~(\ref{4a}) and (\ref{9a}), the Dirac points appear when the pairs $(u_i,u_i^*)$ and $(u_j,u_j^*)$ are different. This result is in accordance with the considerations of Ref.~\cite{yang}.

\begin{table}\label{tab3a}
\caption{\textbf{The ordinary representations of the point group ${\bf C}_{4h}$}.
The last column shows the quasimomentum-component combinations that are transformed according to these representations.}
\begin{tabular}{c|cccccccc|c}
\hline
\hline \\[-2.5mm]
 &$E$&$C_4$&$C_2$&$C_4^3$&$I$&$IC_4$ &$IC_2$&$IC_4^3$& functions \\
 \colrule
$A_+$&$1$&$1$&$1$&$1$&$1$&$1$&$1$&$1$&$p_z^2$, $p_x^2+p_y^2$ \\
$A_-$&$1$&$1$&$1$&$1$&$-1$&$-1$&$-1$&$-1$&$p_z$ \\
$B_{1+}$&$1$&$-1$&$1$&$-1$&$1$&$-1$&$1$&$-1$&$p_x^2-p_y^2$ \\
$B_{1-}$&$1$&$-1$&$1$&$-1$&$-1$&$1$&$-1$&$1$&$(p_x^2-p_y^2)p_z$ \\
$B_{2+}$&$1$&$i$&$-1$&$-i$&$1$&$i$&$-1$&$-i$&$(p_x-ip_y)p_z$ \\
$B_{2-}$&$1$&$i$&$-1$&$-i$&$-1$&$-i$&$1$&$i$&$p_x-ip_y$ \\
$B_{3+}$&$1$&$-i$&$-1$&$i$&$1$&$-i$&$-1$&$i$&$(p_x+ip_y)p_z$ \\
$B_{3-}$&$1$&$-i$&$-1$&$i$&$-1$&$i$&$1$&$-i$&$p_x+ip_y$ \\
\hline \hline
\end{tabular}
\end{table}

Consider now the inversion of the bands $AB_+$ and $AB_-$ or $B_{12+}$ and $B_{12-}$. In both these cases, the pairs $(u_i,u_i^*)$ and $(u_j,u_j^*)$ are the same, but the bands have different parities. For these bands, we find
 \begin{eqnarray}\label{12a}
|t|^2+|u|^2=v_z^2p_z^2+v_{\perp}^2(p_x^2+p_y^2),
\end{eqnarray}
where $v_z$ and $v_{\perp}$ are some real constants. After the inversion  of these bands at the point $\Gamma$, their crossing does not occur, and two quasi-Dirac points appears at $p_z=\pm p_{z0}$,
 \begin{eqnarray}\label{13a}
p_{z0}=\sqrt{\frac{2}{(\alpha_{i}^z+\alpha_{j}^z)}\left(-\Delta- \frac{v_z^2}{(\alpha_{i}^z+\alpha_{j}^z)}\right)}.
\end{eqnarray}
This $p_{z0}$ is determined by the condition that the direct gap between the bands is minimal at this point $\zeta$ of the $p_z$ axis. It follows from Eq.~(\ref{13a}) that the quasi-Dirac points can exist when $\Delta<\Delta_{\rm cr}$ where the critical value $\Delta_{\rm cr}$ of the half-gap is
\begin{eqnarray}\label{14a}
\Delta_{\rm cr}=- \frac{v_z^2}{(\alpha_{i}^z+\alpha_{j}^z)}.
\end{eqnarray}
In the vicinity of $p_z=p_{z0}$, the dispersion has the form of the quasi-Dirac spectrum,
 \begin{eqnarray}\label{15a}
\varepsilon_{i,j}({\bf p})&=&\varepsilon_d +a_z (p_z-p_{z0}) \nonumber  \\
&\pm&\sqrt{\Delta_{\zeta}^2+\tilde v_z^2(p_z-p_{z0})^2
+\tilde v_{\perp}^2(p_x^2+p_y^2)},
\end{eqnarray}
where $\varepsilon_d$ and $a_z$ are given by formulas (\ref{7a}), and
 \begin{eqnarray}\label{16a}
   \Delta_{\zeta}^2\!&=&(\Delta+ \frac{(\alpha_{i}^z+ \alpha_{j}^z)}{2} p_{z0}^2)^2+v_z^2p_{z0}^2=\Delta_{\rm cr}(2\Delta-\Delta_{\rm cr}),
\nonumber  \\
\tilde v_z^2\!&=&v_z^2+\!(\alpha_{i}^z\!+ \!\alpha_{j}^z)^2p_{z0}^2\!+\!(\alpha_{i}^z\!+\! \alpha_{j}^z)(\Delta+\!\frac{\alpha_{i}^z\!+\!\alpha_{j}^z}{2} p_{z0}^2)  \nonumber \\
&=&2(\alpha_{i}^z+ \alpha_{j}^z)(\Delta_{\rm cr}-\Delta),~~~ \\
\tilde v_{\perp}^2\!\!\!&=&v_{\perp}^2+\Delta_{\rm cr}(\alpha_{i}^{\perp}+ \alpha_{j}^{\perp}). \nonumber
\end{eqnarray}
Note that formulas (\ref{15a}), (\ref{16a}) are applicable to the description of real situations when $|\Delta_{\zeta}|$ is noticeably less than the spacing between the bands $i$, $j$ and the other bands at the point $\zeta$. This can occur if $|\Delta_{\rm cr}|$ is small enough (i.e., if $v_z$ is noticeably less than the typical values of the velocity, $10^5-10^6$ cm/s). In this case, we also have $\tilde v_{\perp}\approx v_{\perp}$. When the negative $\Delta$ is of the order of $\Delta_{\rm cr}$, $|\Delta| \gtrsim |\Delta_{\rm cr}|$, dispersion (\ref{15a}) well describes the charge carriers only if the Fermi energy $\varepsilon_F$ is very close to $\Delta_{\zeta}$ or $-\Delta_{\zeta}$, $|\varepsilon_F|-|\Delta_{\zeta}|\ll |\Delta_{\rm cr}|$. Otherwise, it is necessary to take into account corrections to dispersion relation (\ref{15a}). In this case, formulas (\ref{3a}), (\ref{12a}) can be used, setting $\alpha_i^{x}=\alpha_i^{y}= \alpha_j^{x}=\alpha_j^{y}=0$ in them. When $|\Delta|$ is large, $\Delta_{\zeta}^2\approx 2\Delta_{\rm cr}\Delta$ can be sufficiently small, and Eqs.~(\ref{15a}), (\ref{16a}) are valid at $|\varepsilon_F|\ll |\Delta|$, but $|\varepsilon_F|$ is not necessarily close to $|\Delta_{\zeta}|$.

\begin{table}\label{tab4a}
\caption{\textbf{Multiplication table for the spinor representations of the point group ${\bf D}_{4h}$.} The results of the multiplications are the ordinary representations indicated in Supplementary Table V. The basis states for $E'_l$ ($l=1\pm,2\pm$) are multiplied by the factors $u_l$, $u_l^*$ at the rotation through $\pi/2$ angle. These factor are given in the last column; $\epsilon=\exp{(i\pi/4)}$.}
\begin{tabular}{c|c|c|c|c|c}
\hline
\hline \\[-2.5mm]
 &$E'_{1+}$&$E'_{1-}$&$E'_{2+}$&$E'_{2-}$& ($u_l,u_l^*)$ \\
 \colrule
$(E'_{1+})^*$&$A_1^++A_2^+$&$A_1^-+A_2^-$&$B_1^++B_2^+$& $B_1^-+B_2^-$&$(\epsilon,\epsilon^*)$\\
 &$+E^+$&$+E^-$&$+E^+$&$+E^-$& \\
 \colrule
$(E'_{1-})^*$&$A_1^-+A_2^-$&$A_1^++A_2^+$&$B_1^-+B_2^-$&$B_1^++B_2^+$& $(\epsilon,\epsilon^*)$\\
&$+E^-$&$+E^+$&$+E^-$&$+E^+$& \\
 \colrule
$(E'_{2+})^*$&$B_1^++B_2^+$&$B_1^-+B_2^-$&$A_1^++A_2^+$&$A_1^-+A_2^-$& $(-\epsilon,-\epsilon^*)$\\
 &$+E^+$&$+E^-$&$+E^+$&$+E^-$& \\
 \colrule
 $(E'_{2-})^*$&$B_1^-+B_2^-$&$B_1^++B_2^+$&$A_1^-+A_2^-$&$A_1^++A_2^+$& $(-\epsilon,-\epsilon^*)$ \\
 &$+E^-$&$+E^+$&$+E^-$&$+E^+$& \\
\hline \hline
\end{tabular}
\end{table}

\begin{table}[t]\label{tab5a}
\caption{\textbf{The ordinary representations of the point group ${\bf D}_{4h}$.} The quasimomentum-component combinations that are transformed according to these representations are also indicated. Here $E^+$ and $E^-$ are the two-dimensional representations.}
\begin{tabular}{c|c||c|c}
\hline
\hline \\[-2.5mm]
$A_1^+$& $p_z^2$, $p_x^2+p_y^2$&$A_1^-$& $(p_x^2-p_y^2)p_zp_xp_y$ \\
\colrule
$A_2^+$& $(p_x^2-p_y^2)p_z^2p_xp_y$& $A_2^-$& $p_z$ \\
\colrule
$B_1^+$& $p_x^2-p_y^2$&$B_1^-$& $p_xp_yp_z$\\
\colrule
$B_2^+$& $p_xp_y$ &$B_2^-$& $p_z(p_x^2-p_y^2)$  \\
\colrule
$E^+$& $p_zp_x$, $p_zp_y$&$E^-$& $p_x$, $p_y$  \\
\hline \hline
\end{tabular}
\end{table}

\subsubsection{Group ${\bf D}_{4h}$}

For the group ${\bf D}_{4h}$, all the spinor representations $E'_{1+}$, $E'_{1-}$, $E'_{2+}$, $E'_{2-}$ are two-dimensional \cite{bir}, and their direct products are decomposed into the ordinary irreducible representations of this group (Supplementary Table IV). These ordinary representations and quasimomentum combinations, which are transformed according to these representations, are indicated in Supplementary Table V.

With these Supplementary Tables, for the pairs of the bands $E'_{1+}$, $E'_{2+}$ or $E'_{1-}$, $E'_{2-}$, we find
 \begin{eqnarray}\label{17a}
|t|^2\!+|u|^2\!=c_{t1}^2p_x^2p_y^2\!+c_{t2}^2(p_x^2\!-p_y^2)^2\!+ c_{u}^2p_z^2(p_x^2\!+p_y^2),~~
\end{eqnarray}
where $c_{t1}$, $c_{t2}$, $c_{u}$ are some real constants. The inversion of these bands leads to the pair of the Dirac points described by formulas (\ref{5a}), (\ref{7a}), (\ref{8a}), (\ref{10a}), (\ref{11a}).
For the pairs of the bands $E'_{1+}$, $E'_{2-}$ or $E'_{1-}$, $E'_{2+}$, we arrive at
 \begin{eqnarray}\label{18a}
|t|^2\!\!+|u|^2\!=c_{t1}^2p_x^2p_y^2p_z^2\!+\!c_{t2}^2p_z^2(p_x^2\!- \!p_y^2)^2\!\!+ \!v_{\perp}^2(p_x^2\!+p_y^2),~~~
\end{eqnarray}
where $c_{t1}$, $c_{t2}$, $v_{\perp}$ are some real constants. The inversion of these bands produces the pair of the Dirac points described by formulas (\ref{5a})-(\ref{8a}).

On the other hand, for the pairs of the bands $E'_{1+}$, $E'_{1-}$ or $E'_{2+}$, $E'_{2-}$, Supplementary Tables IV and V yield
 \begin{eqnarray}\label{19a}
|t|^2\!\!+|u|^2\!=c_{t}^2p_x^2p_y^2p_z^2(p_x^2\!- \!p_y^2)^2\!\!+\!v_{z}^2p_z^2\!\!+ \!v_{\perp}^2(p_x^2\!+p_y^2),~~
\end{eqnarray}
where $c_{t}$, $v_{z}$, $v_{\perp}$ are some real constants. The inversion of these bands leads to the pair of the quasi-Dirac points described by formulas (\ref{13a})-(\ref{16a}).

Finally, it is worth noting that the point group of $\Gamma$ can be ${\bf O}_{4h}$. However, in this case, the symmetry requires $v_z=v_{\perp}$, $\alpha_{i,j}^z=\alpha_{i,j}^{\perp}$, and the velocity $v_z$ as well as the critical value of the gap defined by Eq.~(\ref{14a}) are not small. In this situation, formulas (\ref{13a})-(\ref{16a}), obtained within the two-band model, can hardly be used to describe the quasi-Dirac spectra near the $\Gamma$ point with the cubic symmetry ${\bf O}_{4h}$.

Thus, for each of the groups ${\bf C}_{4h}$ and ${\bf D}_{4h}$, the inversion of the bands leads to the pair of the Dirac or  quasi-Dirac points in the $4$-fold axis. This statement remains true for other axes, and we expect the quasi-Dirac points are not unusual in crystals. In particular, in the $2$-fold axis, only the pair of the quasi-Dirac points can occur. Indeed, in this case, the group of the $\Gamma$ point is either ${\bf C}_{2h}$ or ${\bf D}_{2h}$. For ${\bf C}_{2h}$ group, its four spinor representations are one-dimensional. When they are combined into the pairs, we get only two bands similar to $AB_+$, $AB_-$. The inversion of these bands produces the quasi-Dirac point which is described by formulas (\ref{12a})-(\ref{16a}). The only difference is that we should use $v_x$, $v_y$ and $\alpha_{i,j}^{x}$, $\alpha_{i,j}^{y}$ instead of   $v_{\perp}$ and $\alpha_{i,j}^{\perp}$, respectively. For ${\bf D}_{2h}$ group (this is the group of the $\Gamma$ point in ZrTe$_5$), there are two two-dimensional representations differing in the parity, and after the band inversion, these representations also lead to a pair of the quasi-Dirac points described by formulas (\ref{12a})-(\ref{16a}).

\subsection{Supplementary Note $2$: Frequencies and cyclotron masses for charge carriers near  quasi-Dirac points}

For the charge carriers with the dispersion $\varepsilon_{c}({\bf p})$ or $\varepsilon_{v}({\bf p})$ defined by \begin{eqnarray}\label{1b}
\varepsilon_{c,v}({\bf p})=\varepsilon_d+{\bf a}{\bf p}\pm \sqrt{\Delta^2+v_{x}^2p_x^2 +v_{y}^2p_y^2+v_{z}^2p_z^2},~~
\end{eqnarray}
the Fermi surface, $\varepsilon_{c,v}({\bf p})=\varepsilon_F$,  has the shape of the ellipsoid defined by the equation:
 \begin{eqnarray}\label{2b}
 v_x^2p_x^2+v_y^2p_y^2+v_z^2(1-\tilde a_z^2)p_z^2=\frac{(\varepsilon_F-\varepsilon_d)^2-\Delta_{\rm min}^2}{(1-\tilde a_z^2)},~~
 \end{eqnarray}
where  ${\bf a}=(0,0,a_z)$, $\tilde a_z\equiv a_z/v_z$, $\Delta_{\rm min}=\Delta (1-\tilde a_z^2)^2$. With equation (\ref{2b}), we find the maximal cross sectional areas $S_{{\rm max},i}$ of the Fermi surface, the frequencies $F_i$ of quantum oscillations,  and the cyclotron masses $m_{*,i}$ of the electron orbits,
 \begin{eqnarray}\label{3b}
 \frac{2 e\hbar F_z}{c}\!&=&\!\frac{S_{{\rm max},z}}{\pi}\!=\! \frac{(\varepsilon_F\!-\!\varepsilon_d)^2\!-\!\Delta_{\rm min}^2}{v_xv_y(1-\tilde a_z^2)}, \\    m_{*,z}\!\!&=&\!\!\frac{(\varepsilon_F\!-\!\varepsilon_d)}{v_xv_y(1-\tilde a_z^2)},~~~~~ \\
 \frac{2 e\hbar F_x}{c}\!&=&\!\frac{S_{{\rm max},x}}{\pi}\!=\! \frac{(\varepsilon_F\!-\!\varepsilon_d)^2\!-\!\Delta_{\rm min}^2}{v_zv_y(1-\tilde a_z^2)^{3/2}}, \\    m_{*,x}\!\!&=&\!\!\frac{(\varepsilon_F\!-\!\varepsilon_d)}{v_zv_y(1-\tilde a_z^2)^{3/2}},~~~~~\label{6b}
  \end{eqnarray}
where the subscript $i=x,y,z$ means that all these quantities correspond to the magnetic field directed along the $i$th axis. Expressions for $F_y$ and $m_{*,y}$ are obtained by the replacement $v_y\to v_x$ in the formulas for $F_x$ and $m_{*,x}$. It follows from Eqs.~(\ref{3b})-(\ref{6b}) that for all $i$, the ratio $F_i/m_{*,i}$ is one and the same,
  \begin{eqnarray}\label{7b}
 \frac{2e\hbar F_i}{c|m_{*,i}|}&=&\frac{(\varepsilon_F - \varepsilon_d)^2-\Delta_{\rm min}^2}{|\varepsilon_F - \varepsilon_d|}\equiv \tilde\varepsilon_F.
 \end{eqnarray}
This property is the hallmark of the quasi-Dirac spectrum.
Formula (\ref{7b}) also means that for any $i$ and $j$,
  \begin{eqnarray} \label{8b}
 \frac{F_i}{F_j}&=&\frac{m_{*,i}}{m_{*,j}} \equiv \tilde \epsilon_{ij},
  \end{eqnarray}
where $\tilde \epsilon_{xy}=\epsilon_{xy}$, $\tilde \epsilon_{zx}/\epsilon_{zx}=\tilde \epsilon_{zy}/\epsilon_{zy}=(1-\tilde a_z)^{1/2}$, and
$\epsilon_{ij}=v_i/v_j$ is the anisotropy of the velocities.
With formulas (\ref{2b}) and (\ref{8b}), we also find the angular dependence of the oscillation frequency $F$ when the magnetic field lies in the $i-j$ plane,
\begin{eqnarray}\label{9b}
F(\theta)=\frac{F_i}{(\cos^2\theta + \tilde\epsilon_{ij}^2 \sin^2\theta)^{1/2}},
\end{eqnarray}
where $\theta$ is the angle between $H$ and the $i$ axis.

\subsection{Supplementary Note $3$: Magneto-optical conductivity for quasi-Dirac points}

Magneto-optical experiments make it possible to find the parameters of quasi-Dirac spectrum (\ref{1b}), including the gap $2\Delta$.
For this spectrum, the explicit form of the Landau subbands $\varepsilon^{(l)}(p_n)$, satisfying the equation
  \begin{equation}
S_{c,v}(\varepsilon^{(l)},p_n)=\frac{2\pi\hbar e H}{c}l,
\end{equation}
looks as follows \cite{m-sh,m-sh19}:
\begin{eqnarray}\label{1c}
\varepsilon_{c,v}^{(l)}(p_n)= \varepsilon_d+v_np_n\pm\sqrt{2\frac{e\hbar Hv_{\perp}^2}{c}l+\Delta^2({\bf n})+L_np_n^2},~~~~~
\end{eqnarray}
where $l=0,1,2,\dots$; $v_n$, $v_{\perp}$, $L_n$, and $\Delta({\bf n})$ are some constants which depend on the direction of the magnetic field, ${\bf n}\equiv {\bf H}/H$;  $p_n$ is the component of the quasimomentum along the magnetic field, and  $S_{c,v}(\varepsilon^{(l)},p_n)$ is the area of the cross-section of the Fermi surface by the plane $p_n=$const. The Landau subbands $\varepsilon_{c}^{(l)}(p_n)$ and $\varepsilon_{v}^{(r)}(p_n)$ in the $c$ and $v$  bands have minima and maxima, respectively, and these maxima and minima are shifted with respect to the point ${\bf p}=0$ (see, e.g., Fig.~1 in the main text). If the Fermi level $\varepsilon_F$ lies between the subbands $\varepsilon_{v}^{(r)}(p_n)$ and $\varepsilon_{c}^{(l)}(p_n)$ where the integer $l$ and $r$ satisfy a certain selection rule \cite{jiang20,mohelsky23}, the magneto-optical conductivity $\sigma(\omega)$ exhibits a peak at the  frequency $\omega$ coinciding with the gap $\varepsilon_{c}^{l}- \varepsilon_{v}^{r}$ at the point $p_n$ defined by the condition, $d[\varepsilon_{c}^{l}(p_n)- \varepsilon_{v}^{r}(p_n)]/dp_n=0$. With Eq.~(\ref{1c}), this condition gives $p_n=0$ and the following expression for $\omega$:
 \begin{equation} \label{2c}
\hbar\omega\!=\!\sqrt{2\frac{e\hbar Hv_{\perp}^2}{c}l+\Delta^2({\bf n})} +\!\!\sqrt{2\frac{e\hbar Hv_{\perp}^2}{c}r+ \Delta^2({\bf n})}.
 \end{equation}
At ${\bf a}=(0,0,a_z)$, the velocity $v_{\perp}$ and the half-gap $\Delta({\bf n})$ are determined by the formulas:
\begin{eqnarray}\label{3c}
\Delta^2({\bf n})&=&\Delta^2 \frac{n_z^2+(1-\tilde a_z^2)(\epsilon^2_{zx}n_x^2+\epsilon^2_{zy}n_y^2)}{ n_z^2+\epsilon^2_{zx}n_x^2+\epsilon^2_{zy}n_y^2}, \\
v_{\perp}^2&=&v_xv_y \frac{[n_z^2+(1-\tilde a_z^2)(\epsilon^2_{zx}n_x^2+\epsilon^2_{zy}n_y^2)]^{3/2}}{ n_z^2+\epsilon^2_{zx}n_x^2+\epsilon^2_{zy}n_y^2}, \label{4c}
\end{eqnarray}
where $\epsilon_{zx}=v_z/v_x$, $\epsilon_{zy}=v_z/v_y$ and $\tilde a_z=a_z/v_z$ define  the anisotropy of the velocities and the tilt of the spectrum, respectively. It is important to emphasize that for the nonzero tilt, the measured gap $2\Delta({\bf n})$ depends on the direction of the magnetic field, and its value lies between the minimal indirect gap $2\Delta_{\rm min}=2\Delta(1-\tilde a_z^2)^{1/2}$ and the minimal direct gap $2\Delta$ of the spectrum without the magnetic field. In particular, at $H\parallel z$, Eqs.~(\ref{3c}) give $\Delta({\bf n})=\Delta$ and $v_{\perp}^2=v_xv_y$, whereas $\Delta({\bf n})=\Delta(1-\tilde a_z^2)^{1/2}$ and  $v_{\perp}^2=v_yv_z(1-\tilde a_z^2)^{3/2}$  if $H\parallel x$. These angular dependences of $\Delta({\bf n})$ and $v_{\perp}$ may be important in analyzing experimental data if the normal to the surface of the sample does not coincides with one of the coordinate axes $x, y, z$. Indeed, in this case, the magnetic field perpendicular or parallel to the surface is inclined to the $z$ axis.

\end{document}